\documentclass{aa}  

\usepackage{graphicx}
\usepackage{txfonts}
\usepackage[colorlinks,linkcolor=blue,anchorcolor=red,citecolor=green]{hyperref}
\usepackage{multirow}

\begin{document} 

   \title{Capture of satellites during planetary encounters}
   \subtitle{A case study of the Neptunian moons Triton and Nereid}

   \author{Daohai Li\inst{1}\fnmsep\thanks{li.daohai@astro.lu.se; lidaohai@gmail.com}\and
          Anders Johansen\inst{1}\and
          Alexander J. Mustill\inst{1}\and
          Melvyn B. Davies\inst{1}
          \and
          Apostolos A. Christou\inst{2}
          }

   \institute{Lund Observatory\\ Department of Astronomy and Theoretical Physics, Lund University\\ Box 43, SE-221 00 Lund, Sweden
         \and
             Armagh Observatory and Planetarium\\ College Hill, Armagh, BT61 9DG \\ Northern Ireland, UK
             }
\authorrunning{Li et al.}


 
  \abstract
   {Single-binary scattering may lead to an   exchange where the single object captures a component of the binary, forming a new binary. This has been well studied in encounters between a star--planet pair and a single star.}
   {Here we explore the application of the exchange mechanism to a planet--satellite pair and another planet in the gravitational potential of a central star. As a case study, we focus on encounters between a satellite-bearing object and Neptune. We investigate whether Neptune can capture satellites from that object and if the captured satellites have orbits analogous to the Neptunian moons Triton and Nereid.}
   {Using $N$-body simulations, we study the capture probability at different encounter distances. Post-capture, we use a simple analytical argument to estimate how the captured orbits evolve under collisional and tidal effects.}
   {We find that the average capture probability reaches $\sim$$10\%$ if Neptune penetrates the donor planet's satellite system. Most moons grabbed by Neptune acquire highly eccentric orbits. Post-capture, around half of those captured, especially those on tight orbits, can be circularised, either by tides only or by collisions+tides, turning into Triton-like objects. Captures further out, on the other hand, stay on wide and eccentric orbits like that of Nereid. Both  moon types  can be captured in the same encounter and they have wide distributions in orbital inclination. Therefore, Triton naturally has a $\sim$50\% chance of being  retrograde.}
   {A similar process potentially applies to an exoplanetary system, and our model predicts that exomoons can jump from one planet to another during planetary scattering. Specifically, there should be two distinct populations of captured moons: one on close-in circular orbits and the other on far-out  eccentric orbits. The two populations may have highly inclined prograde or retrograde orbits.}


   \maketitle
%

\section{Introduction}
\label{sec-intro}

It is well established \citep[e.g.][]{Heggie1975,Hut1983} that when a binary  A+B encounters a third object  C, C may replace one of the binary’s components (A, for instance), forming a new binary C+B. If B is much less massive than either A or C, as is the case in this work, we say that B is captured by C. This has been exemplified by encounters between a star+planet pair and a star \citep{Malmberg2011}.

Here we examine this mechanism on a much smaller scale and in the presence of a fourth body, D. Specifically, we investigate encounters between a planet+satellite pair and another planet in the gravity field of a central star D. The only work we know that touches upon this topic is \citet{Hong2018}. There the authors run sets of two simulations. In the first, they let a moon-hosting planet and a second planet  encounter, and they derived the capture rate of the moons as a function of the encounter distance and the planet masses. In the second simulation, systems of three moon-bearing planets were followed for 100 Myr. The final fate (i.e. whether they were captured)  and orbital distribution of the moons were analysed.

Here we explore satellite capture during planetary encounters in more detail. Our work is not to be regarded as a general study of exoplanetary systems, but rather as a case study. We approach the problem from the context of a dynamical instability period in the early Solar System where the giant planets may have undergone close encounters with each other (see below). More specifically, we focus on encounters between Neptune and another planet, and we investigate whether the Neptunian moons Triton and Nereid can be acquired or captured from  the other planet during these encounters. These two moons have unique orbital characteristics among the Solar System moons and are believed to have been captured by Neptune \citep[e.g.][and see below]{Agnor2006,Nesvorny2007}. We aim to examine whether exchange during planet--planet encounters is a valid alternative origin model. In addition to the capture process {per se}, we also model the post-capture evolution and discuss briefly the applicability of this mechanism to extrasolar planetary systems.

All four giant planets in our Solar System have two types of satellites as classified by their orbital features. The first (the regular satellites) are characterised by close-in, circular, and coplanar orbits and the second (the irregular satellites)  by wide, highly eccentric, and highly inclined trajectories. A quantitative limit between these two classes is where the solar perturbation is as important as that of the host planet's oblateness \citep{Burns1986a,Li2016}. The regulars formed in situ in the circumplanetary disc \citep[e.g.][]{Canup2002,Mosqueira2003a}, while the irregulars were captured from heliocentric orbits, for example, via three-body interaction \citep{Nesvorny2007}.

However,  not all moons fit in this classification. Uniquely in the Solar System, Triton orbits its host planet, Neptune, on a close-in circular yet retrograde path \citep{Murray1999}. However,  according to the above definition, it is a regular moon; the obvious inconsistency between its retrograde orbit and accretion in the circum-Neptunian disc  strongly suggests that it was indeed captured by Neptune.

Most likely, Triton's capture happened in the early Solar System. To set up the context, we first briefly review the Nice scenario, a widely accepted early Solar System evolution model. There, the giant planets, owing to the interaction with a massive planetesimal disc, radially migrated long distances. A key feature of it is an instability period when a global upheaval of the entire Solar System occurs with giant planets gaining significant eccentricities and encountering each other at close distances. In its original version \citep[e.g.][]{Tsiganis2005}  these encounters usually occur between Saturn, Uranus, and Neptune \citep{Nesvorny2007,Morbidelli2009}. Then, a recent variant, known as the jumping Jupiter model, showed that a Solar System starting with the two gas giants and three ice giants (IGs; the additional IG was ejected during the instability period) could better meet the observational constraints \citep[][and cf. \citealt{Morbidelli2009,Brasser2009}]{Nesvorny2011, Batygin2012a,Nesvorny2012}. In this version, all giant planets seem to participate in close encounters \citep{Nesvorny2014}, probably all involving the same IG that is ejected later.

Now we turn back to the capture of Triton. A promising mechanism is the exchange capture model \citep{Agnor2006}, postulating that when an asteroid binary encounters Neptune at a low relative velocity, one of the components may acquire a bound orbit around the planet. The chance to procure a Triton-mass object, examined in the Nice scenario, is 2\%-50\% \citep{Vokrouhlicky2008,Nogueira2011}. In addition, this model may actually precede the Nice scenario, featuring a higher efficiency \citep{Vokrouhlicky2008}.

Nonetheless, the current orbital feature of Triton does not necessarily preclude an  in situ formation scenario and its retrograde orbit might be acquired later. \cite{Harrington1979} suggested that a rogue planet might flip Triton's orbit when it encountered Neptune. Recently, \cite{Li2020} proposed a similar model embedded in the Nice scenario and held the then-ejected IG responsible for Triton's orbital inclination.

The peculiarity of the Neptunian moon system does not stop at Triton. As discussed in \citet{Li2020}, the irregular satellite Nereid has the highest eccentricity among all Solar System moons and a relatively tight orbit deep inside Neptune's Hill radius (compared to other irregulars). If captured from the primordial planetesimal disc \citep{Nesvorny2007}, Nereid would be the largest object captured by the giant planets (among Trojans and irregular satellites) from this reservoir during the instability period of the Nice scenario. However, the low capture efficiency for irregular satellites \citep{Nesvorny2014} combined with the sparseness of large, Nereid-sized \citep[$\sim$170 km in radius;][]{Murray1999} objects in the reservoir \citep[e.g.][]{Nesvorny2016} leads to a probability of 0.6\% for its procurement at Neptune \citep{Li2020}. It has been proposed that Nereid  was also  a regular moon formed in the circum-Neptunian disc, which was  then scattered onto its current orbit by Triton \citep{Goldreich1989} or by another planet \citep{Harrington1979,Li2020}.

The planetary close encounters predicted in the Nice scenario give rise to rich dynamics. During these events, like those discussed above, nearby heliocentric planetesimals may be captured by Neptune as irregular satellites due to three-body interactions \citep{Nesvorny2007,Nesvorny2014}. In addition, Trojan objects can be acquired as a result of the instantaneous displacement of the planet's orbit and thus its L4 and L5 libration regions \citep{Nesvorny2013}. In a similar sense, if Neptune's semimajor axis changes by a significant fraction of an au in a single encounter when it is  approximately 28 au, objects previously trapped in mean motion resonance with Neptune can be dropped, forming the so-called kernel \citep{Petit2011} of the Kuiper belt \citep{Nesvorny2015,Deienno2017}, a pileup of orbits at $\sim$$44$ au with low eccentricities and inclinations. \footnote{Alternatively, the kernel could represent the edge of the planetesimal disc left mildly perturbed \citep{Gomes2018}, thus removing the requirement that Neptune has an instability at 28 au.} Then, as mentioned already, \citet{Li2020} showed that both Triton and Nereid, if formed in the circum-Neptunian disc, could be simultaneously put on evolutionary paths towards their observed orbits by the encounter \citep[see also][]{Harrington1979}.

We note that in the original Nice scenario, only the four extant giant planets were invoked. This means that a Neptune-encountering planet has to be either Jupiter, Saturn, or Uranus. This may be problematic because these three planets all have cold satellite systems that would have been excited and/or disrupted during the encounter if deep enough \citep{Deienno2014,Nesvorny2014a,Li2020}. In contrast, the jumping Jupiter model predicts that such encounters only occur between Neptune and the then-ejected IG. While the former has an {unusual} satellite system, possibly a result of the encounter, the latter has been ejected and thus presents no constraints. So, the effects reviewed in the previous paragraph, as well as another to be investigated in this paper, are likely only relevant in this jumping Jupiter model with three IGs. We  discuss this further in Section \ref{sec-dis}.

As explained at the beginning of this section, we are inspired by simulations of planet captures during stellar encounters. We want to answer the following questions in this work: Can a satellite be captured from the IG by Neptune during their encounter? What is the capture efficiency?  and What are the orbital characteristics of captured satellites? 

We have organised the paper as follows. The setup of our numerical simulation is described in Section \ref{sec-sim}. The methods used to analyse and classify the captured moons are presented in Section \ref{sec-class}. We then present the results in Sections \ref{sec-res-IG18}, \ref{sec-res-IG10}, and \ref{sec-primo-irre}, showing that both Triton's and Nereid's observed orbits can be created by our mechanism. In Section \ref{sec-dis} we derive the efficiency of our model and its generalisation to other systems. We conclude the paper with Section \ref{sec-con}.

\section{Capture simulations}\label{sec-sim}
Our IG--Neptune encounter assumption is based on the Nice scenario \citep{Tsiganis2005,Nesvorny2012}. Ideally, we wanted like to run the full simulations of Nice scenario and to use the IG--Neptune encounters recorded therein. However, the success rate of such simulations is low \citep{Nesvorny2012} and the encounter statistics are highly stochastic, varying from run to run \citep{Nesvorny2014a,Deienno2014}. Hence, it was uneconomic to simulate the full Nice scenario as we only needed one deep encounter. Therefore, we decided to create the IG--Neptune encounters.

\subsection{Creation of the IG--Neptune encounters}

We use the same method as \cite{Li2020} to generate the encounters  \citep[see also][]{Cloutier2015} in a three-body system of Sun-IG-Neptune. In brief, this is done through generating random position and velocity vectors for the two planets' centre of mass with respect to the Sun and the relative position and velocity vectors between the two planets at their closest approach $q_\mathrm{enc}$. Then the three-body system is propagated backwards and forwards, each for 10 years, and we noted   the heliocentric orbits of Neptune at the end of the two simulations. If the Neptunian orbits are compatible with the constraints from observations of the Kuiper belt \citep[e.g. on the cold classicals; ][]{Dawson2012} and the full Nice scenario simulations \citep{Nesvorny2018}, for example, if its eccentricity and inclination are not high, we deem the encounter realistic and keep it, otherwise it is abandoned. \footnote{We also note that a more violent instability is not necessarily inconsistent with the cold classicals \citep{Gomes2018}.} In this way, we gain complete control over $q_\mathrm{enc}$. For full details we refer to \citet{Li2020}. All $N$-body simulations described in this work were carried out using the {\small MERCURY} code \citep{Chambers1999}.

In \cite{Li2020}, we obtained 500 realisations of IG--Neptune encounters at $q_\mathrm{enc}=0.003$ au \cite[the deepest presented in the literature as observed in full Nice scenario simulations; see][]{Nesvorny2014a}. Additionally, for this work, we have newly generated encounters with $q_\mathrm{enc}$, increasing in a geometric manner, of 0.006, 0.012, 0.024, 0.048, 0.096, 0.192, and 0.384 au, each of 500 realisations. The eight distances correspond to 18.2, 36.5, 72.9, 145.8, 292, 583, 1167, and 2333 times the Neptunian radii ($R_\mathrm{Nep}).$ We note Neptune's current Hill radius (within which the planet's gravity dominates over that of the Sun) is $\sim0.8$ au or $\sim 5000R_\mathrm{Nep}$.

The mass of the IG is not stringently constrained, but it is likely $\gtrsim$ 10 Earth masses \citep[e.g.][]{Morbidelli2009}. Here we introduce two masses for our IG. Firstly, we adopt a value of 18 Earth masses \citep{Nesvorny2014a}, close to that of Neptune, and we assume that the radius of the IG is the same as Neptune's, $1R_\mathrm{Nep}$. Secondly, we introduce a lower IG mass of 10 Earth masses. To create a primordial satellite population around it (see below), we need to derive a radius for it. We have tried the empirical relations for Solar System planets \citep{Lissauer2011} and for the exoplanets \citep{Weiss2014}. The two resulting radii differ from that of Neptune's by at most 20\%. Given that our satellite distribution has already covered a wide range (see below), we just let the IG's radius be $1R_\mathrm{Nep}$ for consistency and simplicity. So we have two models for the IG: one of 18 Earth masses (IG18) and the other of 10 (IG10); both have the same radius as that of Neptune, $1R_\mathrm{Nep}$.

\subsection{Distribution of primordial satellites}
As discussed in Section \ref{sec-intro},  two categories of moons currently coexist around the giant planets. The inner regular moons are primordial, and thus  formed before the instability period. The irregulars, on the other hand, may or may not have been captured before the encounter.

Here for the primordial inner regular satellite region we generate 1100 moons at the IG as massless test particles on circular orbits evenly distributed between 5 to 60 planetary radii, roughly where the Solar System major moons are (e.g. Io at Jupiter and Iapetus at Saturn are at 6 and 61 planetary radii, respectively). In au, the inner and outer boundaries are 0.00082 and 0.099 au, respectively.

These inner regular moons should have accreted on the local Laplace plane of the IG, inclined with respect to the planetary orbital plane by an angle called obliquity. The IG may have gained a large obliquity by a giant collision \citep[e.g.][]{Morbidelli2012b}, predating the encounter or its spin-orbit misalignment may remain small. On the other hand, the encounters are roughly randomly distributed in the solid angle \citep{Li2020}, suggesting that from a statistical point of view, the direction of the IG's spin axis is unimportant. Hence, for simplicity, we assume that the IG's equator is parallel to the {invariant plane}, to which the normal is aligned with the total orbital angular momentum of the Sun-IG-Neptune system.

In Section \ref{sec-intro}, we introduced various models for the origin of irregular satellites, some applicable to the late growth stage of the host planet \citep{Pollack1979} and some to its later evolution \citep{Nesvorny2007}. Both mechanisms could precede the encounters we examine.

The orbit distribution of primordial irregular satellites might not be the same as for the currently observed population due to planet--planet encounters \citep{Li2017a}. Here we create 400 test moons evenly distributed between $\in(100R_\mathrm{Nep},900R_\mathrm{Nep})$, i.e. within 0.016 and 0.15 au. These limits are somewhat arbitrary, with the inner bound chosen to render these outer satellites distinct from the inner regular moons. The outer bound is smaller than the observed population, but is close to that assumed in \citet{Beauge2002}, where the authors studied the effect of planetary migration on these outer moons. Moreover,   for simplicity  we   assume circular orbits \citep{Beauge2002} for our primordial irregular satellites as capture efficiency is not a strong function of the initial eccentricity \citep[e.g.][]{Hut1983,Jilkova2016}, and we  also place them on the invariant plane, different from the extant population but compatible with \citet{Beauge2002}.

Hereafter we refer to the region $\in(5R_\mathrm{Nep},60R_\mathrm{Nep})$ containing the inner moons as the inner region and the region $\in(100R_\mathrm{Nep},900R_\mathrm{Nep})$ containing the outer moons as the outer region, respectively.

\subsection{Capture of satellites  during the encounters}
We then add these test moons to the Sun-IG-Neptune three-body system, as in each of the created encounters, and track their evolution. For those in the inner moon region, capture is only possible at small encounter distances, so we restrict our simulation to $q_\mathrm{enc}\le0.024$ au; for the outer moon region, all eight encounter distances are tested. For capture from the inner moon region both IG18 and IG10 are examined, while for the outer region only IG18 is investigated. Our simulation setup is summarised in Table \ref{tab-sim-sum}.

\setlength{\tabcolsep}{0.4em} 
\begin{table}[h]
\caption{Summary of the different sets of simulations. From left to right, we
present the radial locations of the test moons in $R_\mathrm{Nep}$, the mass of the IG in Earth masses, the encounter
distance in au, and the section in this paper where the results are presented.}
\begin{center}
\label{tab-sim-sum}
\begin{tabular}{cccc}
\hline
\hline
Region ($R_\mathrm{Nep}$)&IG ($m_\oplus$)&$q_\mathrm{enc}$ (au)&Sect.\\
\hline
inner$\in$(5,60)&IG18 (18)&0.003,0.006,0.012,0.024&\ref{sec-res-IG18}\\
\hline
inner$\in$(5,60)&IG10 (10)&0.003,0.006,0.012,0.024&\ref{sec-res-IG10}\\
\hline
\multirow{ 2}{*}{outer$\in$(100,900)}&\multirow{ 2}{*}{IG18 (18)}&0.003,0.006,0.012,0.024&\multirow{ 2}{*}{\ref{sec-primo-irre}}\\
&&0.048,0.096,0.192,0.384&\\
\hline
\end{tabular}
\end{center}
\end{table}

During an encounter, a small fraction of the moons collide with either the IG or Neptune, and are removed from the simulation. Post-encounter, we output the state vectors of the planets and satellites and calculate the planetocentric orbits of the moons. If a moon's Neptunian-centric orbit is eccentric and its semimajor axis $a$ is smaller than half of Neptune's Hill radius \citep{Nesvorny2003}, we recognise it as a (temporary) capture by Neptune.

Not all these captured moons are stable. Those on wide orbits may be subject to large-amplitude Kozai-Lidov cycles \citep{Kozai1962,Lidov1962}. For these moons, we determine the range of the variation of their eccentricities $e$ using the Kozai--Lidov formulation, and eliminate those orbits extending out of Neptune's Hill radius or reaching down to the surface of Neptune. The result is a removal of moons with inclinations close to $90^\circ$. For the inner moons, this effect is suppressed by Neptune's oblateness \citep{Burns1986a,Li2016} and they are thus free from this elimination.

We consider all remaining Neptune-bound objects as captured moons and only consider them in the following.

\section{Classification of the captured moons: two orbital circularisation models} \label{sec-class}

\begin{figure}
\centering
\includegraphics[width=\hsize]{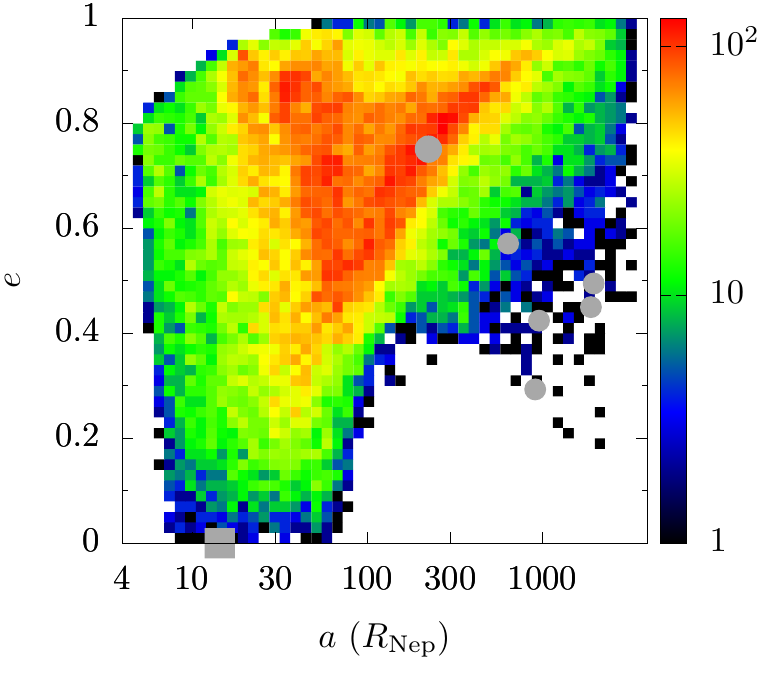}
\caption{ Distribution of $(a,e)$ of the orbits of the captured moons from the inner moon region of IG18 at the encounter distance of $q_\mathrm{enc}=0.003$ au. Warmer colours mean higher numbers of moons. The grey square is Triton; grey circles show the observed irregular satellite population at Neptune, the large one being Nereid; data are taken from Scott S. Sheppard \url{https://sites.google.com/carnegiescience.edu/sheppard/moons/neptunemoons}.}
\label{fig-ae-moon-1-dist}
\end{figure}

In Figure \ref{fig-ae-moon-1-dist}, We show the distribution of the captured moons at Neptune from the inner moon region of IG18 at the encounter distance $q_\mathrm{enc}=0.003$ au. The orbits have wide spreads with $a\in(4R_\mathrm{Nep},3000 R_\mathrm{Nep})$ and $e\in(0,1)$. We note that plotted here are the orbits immediately after the capture, while in the ensuing evolution the orbits may change further. We now describe how we classify the captured moons based on their subsequent orbital evolution.

\begin{figure}
\centering
\includegraphics[width=\hsize]{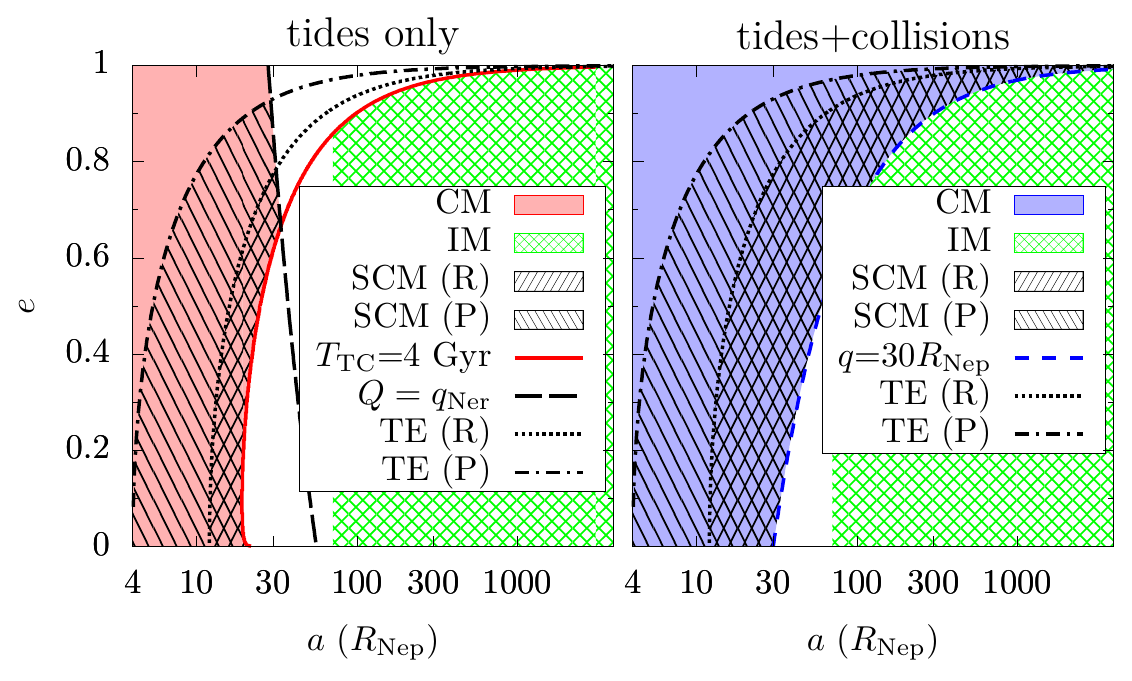}
\caption{Classification of captured moons at Neptune for the two circularisation models. In the left panel we show the tides-only model. The red solid line shows a tidal circularisation isochrone of 4 Gyr; orbits to the left of the line are circularised by tides over the age of the Solar System. The long-dashed line marks a constant apocentre distance equal to Nereid’s pericentre $Q=q_\mathrm{Ner}$, thus moons that lie to the right of the line may collide with Nereid. A circularisable moon (CM) has, by definition, a tidal circularisation time of $<$4 Gyr and, at the same time, does not endanger Nereid (red-shaded region). Due to tidal evolution, a prograde moon to the left of the dash-dotted line [TE (P)] or a retrograde moon to the left of the dotted line [TE (R)] are subject to tidal engulfment by Neptune within 4 Gyr. Correspondingly, moons in the back-slashed (\textbackslash\textbackslash\textbackslash) region are circularisable and survive tidal engulfment on either prograde [SCM (P)] or retrograde orbits [SCM (R);  forward-slashed (\slash\slash\slash) region]. The green-checked region represents irregular moons (IM) with $a>70R_\mathrm{Nep}$ and uncircularisable orbits. In the right panel we show the tides+collisions model. The blue short-dashed line marks a constant pericentre distance $q=30R_\mathrm{Nep}$; orbits on the left (blue-shaded region) can be circularised while those on the right cannot; because collisional evolution is fast, Nereid is essentially safe from collisions and not explicitly considered here. Then TE (P), TE (R), SCM (P), SCM (R), and IM are defined in the same way as in the tides-only model.}
\label{fig-ae-illu}
\end{figure}

Tidal effects, which  mainly involve the distortion in the moon, gradually circularise a moon's orbit \citep{Hut1981}. In Appendix \ref{sec-tid} we outline an analytical model simplified from \citet{Correia2009} to depict this process. Assuming the same tidal parameters as for Triton \citep{Correia2009}, we calculate the tidal circularisation time of each captured moon. This depends sensitively on the size of the orbit \eqref{eq-e-2}. As a result, moons on close-in orbits will be circularised within the age of the  Solar System (4 Gyr) while those on far-out orbits will not. In the left panel of Figure \ref{fig-ae-illu}, the red solid line shows this limit;  any moon on it has the same tidal circularisation time of 4 Gyr.

It has been realised that should the orbit of Triton (or any massive moon) intersect that of Nereid, the latter will be destablised in $10^4-10^6$ yr \citep{Cuk2005,Nogueira2011,Rufu2017,Li2020}. Nonetheless, the tides operate on much longer timescales of hundreds of Myr or Gyr \citep[see e.g. ][]{Goldreich1989,Correia2009,Nogueira2011,Li2020}. For a massive moon to be circularisable by tides only in the presence of Nereid, it must not endanger Nereid (i.e. their orbits should not cross). In the left panel of Figure \ref{fig-ae-illu} the long-dashed line indicates the limit.  We call these circularisable within 4 Gyr by tides and not damaging Nereid {circularisable moons}. This is our first circularisation model: {tides only}. The red-shaded region in the left panel of Figure \ref{fig-ae-illu} shows where these moons are.

The picture changes if Neptune has a primordial satellite system. In this case collisions between Triton and these original moons may efficiently decouple Triton's orbit from Nereid's \citep[cf.][]{Cuk2005}. This process is usually stochastic, and successful orbital decoupling is not guaranteed. For example, if a single preexisting Neptunian moon is as massive as a few per cent of the captured moon, the latter may be disrupted due to the high  collision velocity \citep{Rufu2017}. {Furthermore}, \citet{Li2020} showed that a few tens of small moons totalling only a few per cent of Triton were typically sufficient for orbital decoupling. As such, modelling these collisions obviously depends on the satellite configuration at Neptune, which is unknown and has been highly perturbed by this IG encounter \citep{Li2020}. So here, without performing any actual collisional simulations, we simply assume that the primordial Neptunian moons extend to $30 R_\mathrm{Nep}$, and that any captured moon with a pericentre distance $q<30 R_\mathrm{Nep}$ will collide with the primordial moons and will have its orbit shrunk quickly inside that of Nereid. The  tides will  then gradually circularise the orbit after the depletion of the primordial satellites \citep{Li2020}. This is our second circularisation model: {collisions+tides}. The blue-shaded region in the right panel of Figure \ref{fig-ae-illu} shows the location of these moons.

In the discussion above, the moon's orbital inclination is not considered. This quantity is often measured against its local Laplace plane with respect to which the orbit precesses evenly with time \citep[e.g.][]{Tremaine2009}. For distant moons this plane is close to the orbital plane of the Sun's relative motion with respect to the host planet, while for close-in moons it coincides with the equatorial plane of the host \citep[cf. the definition of irregular and regular satellites,][and see Section \ref{sec-intro}]{Burns1986a}. The current obliquity of Neptune is about $30^\circ$. However, it is unknown whether this obliquity   developed primordially \citep{Morbidelli2012b} or during Neptune's migration and possibly after the capture of the moons  studied here \citep{Boue2010}. Hence, we are unable to define the Laplace plane in our simulations. For simplicity, all moon inclinations are measured against the invariant plane of the Sun-IG-Neptune system. While an appropriate reference frame for moons captured in far-out orbits, it is not ideal for the inner ones. Thus, we only briefly touch on the inclination distribution of our captured moons.

A circularisable moon's orbital evolution does not stop at circularisation. Tides, now primarily related to the deformation in Neptune, make the moon drift inwards or outwards. As a result, the moon may be engulfed by Neptune within the age of the Solar System \citep[e.g.][]{Chyba1989,Correia2009}. In Appendix \ref{sec-tid} we briefly outline a model to determine which moons are subject to this engulfment. It turns out that the speed and direction of tidal drift depend on the angle between Neptune's spin axis and the normal to the orbital plane of the moon. This angle is, however, not defined in our simulations because the spin axis of Neptune is unknown. So we make a simplification by looking into only two situations: (1) all moons are prograde, in the sense that the above-mentioned angle is acute, and (2) all moons are retrograde where the angle is obtuse (Appendix \ref{sec-tid}). When assuming retrograde orbits, the drift of the moons is inwards and faster, hence those on wider orbits may be affected. The dotted lines in the two panels of Figure \ref{fig-ae-illu}  delimit the tidal engulfment for retrograde orbits (those on the left are lost), hence those in the forward-slashed (\slash\slash\slash) region can be circularised (by tides only for the left panel and by tides+collisions for the right panel) and, if retrograde, they survive tidal engulfment. When adopting prograde orbits the limit moves inwards, allowing those on tighter orbits to survive;  the dash-dotted line is used now,  hence those in the back-slashed (\textbackslash\textbackslash\textbackslash) region can be circularised (by tides only for the left panel and by tides+collisions for the right panel) and, if prograde, they survive tidal engulfment. This is derived from our simplified model;  the reality is  less extreme and there should be a spectrum of the inclination for the captures (cf. Figure \ref{fig-enc_inc_rate_sort} below). In this sense, our tidal engulfment estimate can be understood as the upper and lower limits.

All circularisable moons, either through tides only or collisions+tides, are subject to this tidal drift. We refer to the moons that are circularisable and not lost as survivable circularisable moons.

As shown above, only a fraction of the captured moons can be circularised; then the complementary set is uncircularisable. Finding it not straightforward to correlate them with the observed population, we add another criterion $a>70 R_\mathrm{Nep}$ \citep[such that the solar perturbation dominates over the Neptunian $J_2$; e.g.][]{Nogueira2011}. We call these captures {irregular moons}. We recall that tidal (un)circularisation depends on the size of the moon (cf. Equation \ref{eq-e}), and smaller moons are much harder to circularise than Triton (whose the tidal parameters are adopted). So we are probably discussing a lower limit of the irregular moon population if the captured moons are smaller (which is true for the observed irregulars). And  since Nereid is protected in both circularisation models, so are these.

In all figures in this work (except the histograms in Figures \ref{fig-ae-moon-1-dist}, \ref{fig-ae-moon-4-dist}, and \ref{fig-ae-primo-irre}), the   colour red is always associated with (survivable) circularisable moons in the tides-only model, blue with those in the tides+collisions model, and green with irregular moons in either model.

\section{Captured moons at Neptune from the inner moon region around IG18}\label{sec-res-IG18}
Here we focus on our first simulation set:  Neptune's capture of moons from the inner moon region of IG18. The statistics for capture from IG10 and from the outer moon region are discussed in the following two sections.
\subsection{Loss and capture in general}\label{sec-cap-gen}

In Table \ref{tab-cap-num-IG18}, we list the fractions of moons lost from the IG and captured by Neptune in  Cols. 2 and 3 against the encounter distances, listed in the Col. 1. During our closest encounters at $q_\mathrm{enc}=0.003$ au, about half of the moons are lost from the IG, consistent with \citet{Li2020}. Among those lost, roughly 1/5  are captured by Neptune. We recall that as mentioned before, more satellites are initially captured  onto unstable orbits and are not considered here.

\begin{table*}[h]
\caption{Fraction of satellites lost from the inner moon region of IG18 and captured onto different types of orbits around Neptune at different encounter distances. Initially the ice giant (IG18) has $1100\times500=5.5\times10^5$ moons in the inner moon region; after its encounter with Neptune (encounter distances shown in Col. 1), a fraction are lost from the IG (Col. 2). Some are captured by Neptune (Col. 3). Going into the detailed classification, a fraction can be orbitally circularised in 4 Gyr, also not disrupting Nereid, referred to as circularisable moons (CM). However, some of the CMs may be lost to Neptune due to tidal drift; those not absorbed by Neptune are called survivable CMs (SCMs). Moons on retrograde (R) orbits usually drift inwards faster and are thus more likely to be engulfed than those prograde (P). Hence, the fraction of SCMs~(P), where we assume all moons are prograde, is larger than that of SCMs~(R). Uncircularisable moons, if on orbits wider than $70R_\mathrm{Nep}$, are classified as irregular moons (IMs). We have two models for orbital circularisation: tides only and collisions+tides. So we list the fractions of CMs, SCMs (P), SCMs (R), and IMs for the tides-only model in Cols. 4 to 7, and those for the collisions+tides model in Cols. 8 to 11.  All fractions are the number of each type of moons divided by the initial total number at IG18, $5.5\times10^5$. See Figure \ref{fig-ae-illu} for the orbits of each type of  moon.}
\begin{center}
\label{tab-cap-num-IG18}
\begin{tabular}{ccc|cccc|cccc}
\hline
\hline
 \multirow{ 2}{*}{$q_\mathrm{enc}$ (au)}&\multirow{ 2}{*}{loss (\%)}& \multirow{ 2}{*}{capture (\%)} & \multicolumn{4}{c|}{tides only (\%)} & \multicolumn{4}{c}{tides+collisions (\%)}\\
&&&CM&SCM (P)& SCM (R) &IM& CM&SCM (P)& SCM (R) &IM\\
\hline
0.003&52.40&9.46&1.87&1.58&0.53&4.29&5.55&5.19&3.67&3.47\\
0.006&32.06&6.35&1.50&1.29&0.52&2.24&4.23&3.96&2.79&1.77\\
0.012&8.57&3.57&0.58&0.53&0.21&0.94&2.47&2.37&1.81&0.67\\
0.024&0.05&0.05&0&0&0&0.01&0.02&0.02&0.02&0.01\\
\hline
\end{tabular}
\end{center}
\end{table*}

For $q_\mathrm{enc}=0.006$ au, both loss and capture rates drop by $\sim$$1/3$, and the  ratio of 5 roughly holds. For more distant encounters, the numbers of lost and captured satellites decrease dramatically. At $q_\mathrm{enc}=0.024$ au the capture rate is only $0.05\%$. Thus, we believe Neptune cannot capture at larger encounter distances from the inner moon region of IG18. Noticeably, the capture-to-loss ratio increases significantly for these further encounters. For example, it becomes 42\% for $q_\mathrm{enc}=0.012$ au and 93\% at $0.024$ au. This agrees with the case of planet capture during stellar encounters, where during a distant flyby, all planets lost (if any) from one star are captured by the other \citep{Li2019}. Because the number of moons captured during encounters with $q_\mathrm{enc}=0.024$ au is limited, we do not analyse them in detail.

\begin{figure}
\centering
\includegraphics[width=\hsize]{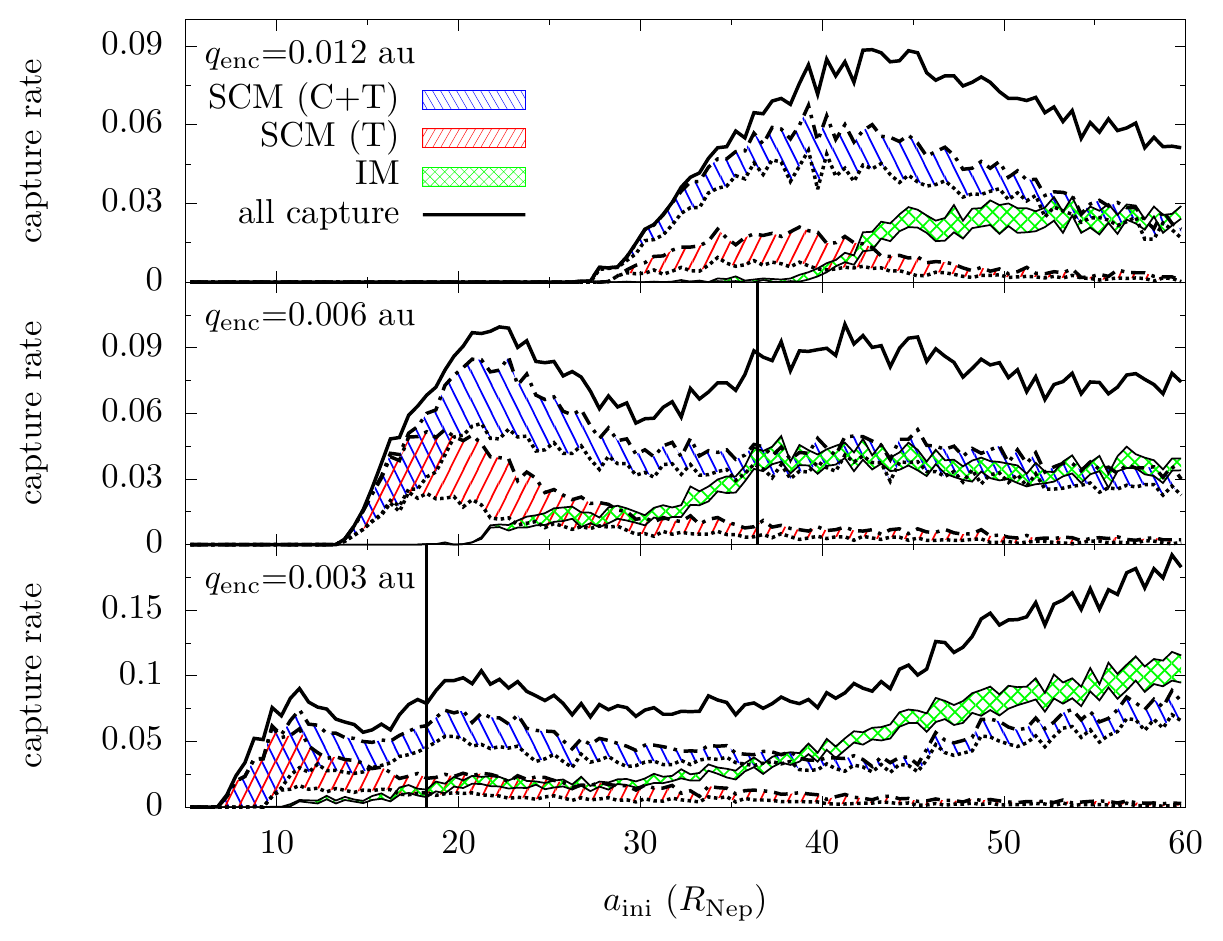}
\caption{Capture rate as a function of initial semimajor axis $a_\mathrm{ini}$ from the inner moon region of IG18 for encounters at different distances. Encounter distances are given in the top left corner of each panel and the vertical lines show the distances in $R_\mathrm{Nep}$. The black thick solid lines are the rates for all captures (Table \ref{tab-cap-num-IG18}, Col. 3, when integrated over $a_\mathrm{ini}$). The red slashed regions indicate those for survivable circularisable moons through tides [SCM~(T)] while keeping Nereid safe, lower bound by tidal engulfment assuming retrograde (Table \ref{tab-cap-num-IG18}, Col. 6) and upper bound by prograde orbits (Col. 5). The blue slashed regions show the efficiencies for the acquirement of survivable circularisable moons though collisions+tides [SCM~(C+T)] and the lower and upper bounds have the same meanings (Table \ref{tab-cap-num-IG18}, Cols. 9 and 10). The green checked regions present capture rates of irregular moons (IM), those uncircularisable and $a>70 R_\mathrm{Nep}$, upper bound by the circularisation model of tides only and lower by collisions+tides (Table \ref{tab-cap-num-IG18}, Cols. 7 and 11). See Figure \ref{fig-ae-illu} for the orbits of each type of the moon.}
\label{fig-a_cap_rate}
\end{figure}

The capture efficiency is  a function of the initial location of the moon $a_\mathrm{ini}$ (around the IG). In Figure \ref{fig-a_cap_rate}, we show this efficiency, defined as the number of moons captured from a given initial $a_\mathrm{ini}$ range divided by that of all moons in that range. Those for $q_\mathrm{enc}=$ 0.003, 0.006, and 0.012 au are represented as the black thick solid lines in the three panels.  In general, the capture efficiency around 10\% and capture starts at $a_\mathrm{ini}\gtrsim q_\mathrm{enc}/3$ \citep[cf.][]{Malmberg2011,Li2019}. Hence, moons inner to $5R_\mathrm{Nep}$ do not need to be modelled for the encounters examined here. While generally the further a satellite is from the IG, the more likely it will be captured by Neptune; however,  the exact dependence on $a_\mathrm{ini}$ is complicated and two local maxima appear:  one just outside $q_\mathrm{enc}/2$ and the other just wide of $q_\mathrm{enc}$. Moreover, this dependence, when scaled over $q_\mathrm{enc}$, is similar for the three encounter distances, implying that it is the ratio$a_\mathrm{ini}/q_\mathrm{enc}$ of a moon that matters.

\begin{figure}
\centering
\includegraphics[width=\hsize]{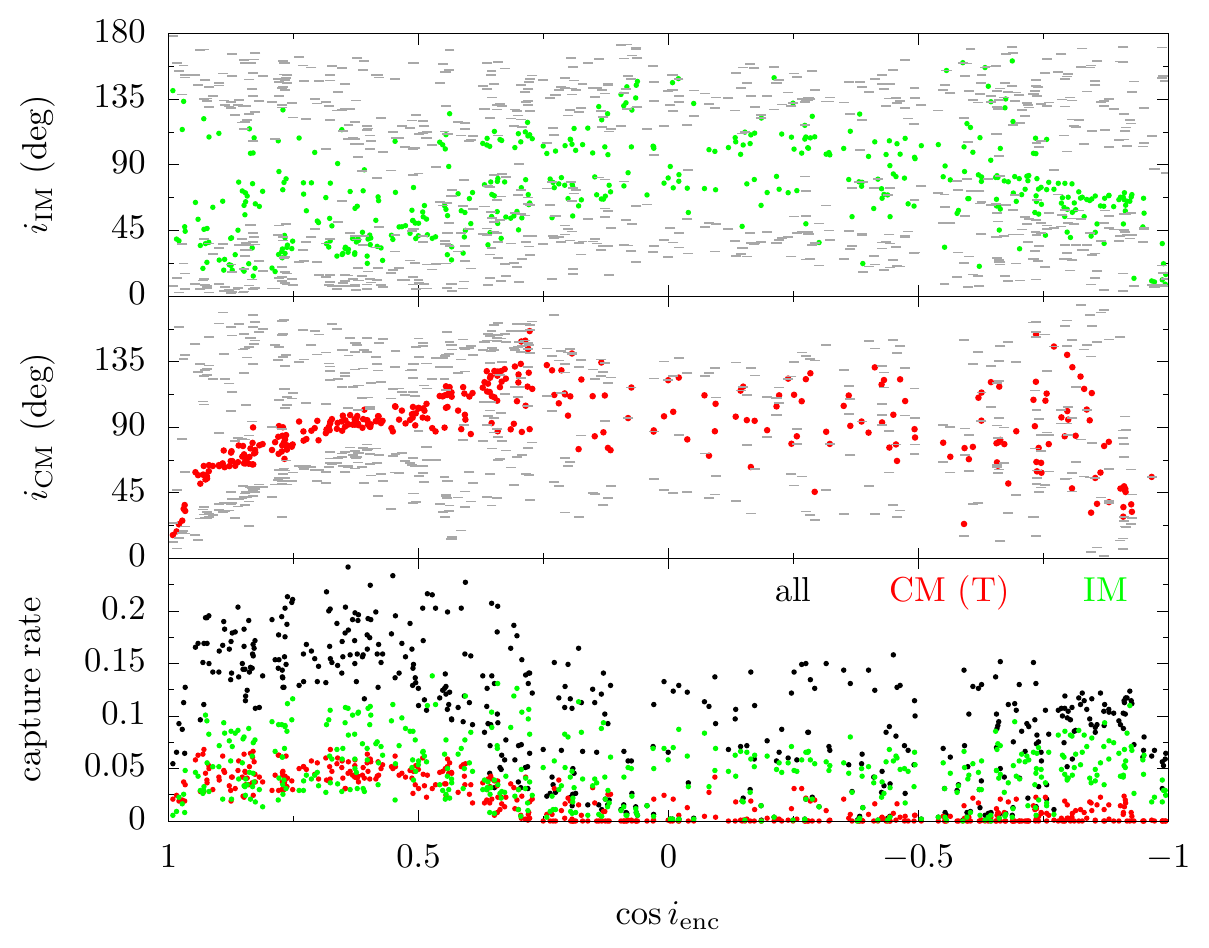}
\caption{Inclination of captured moons and rate of capture for encounters of $q_\mathrm{enc}=0.003$ au from the inner moon region around IG18. The bottom panel shows the capture rate for each encounter as a function of the inclination of that encounter: black dots for all captures (when integrated, this is the first entry of  Table \ref{tab-cap-num-IG18}, Col. 3), red for circularisable moons through tides-only model, CM~(T) (Col. 4), and green for irregular moons (IM, Col. 7). The middle panel shows the inclination distribution of the CM~(T) captured during each encounter: coloured for the median and the lower and upper horizontal line segments the 5th and 95th percentiles. The top panel is the same as the middle, but for IM.}
\label{fig-enc_inc_rate_sort}
\end{figure}

The inclination of the IG's orbit plays an important role in constraining the relative velocity between the IG and the satellites, thus affecting the result of the encounter. In the bottom panel of Figure \ref{fig-enc_inc_rate_sort}, we show the capture efficiency  for each individual encounter as a function of its inclination $\cos i_\mathrm{enc}$ for encounter distance $q_\mathrm{enc}=0.003$ au. From the plot, we see that prograde encounters can capture more efficiently \citep[see also][]{Jilkova2016,Li2019} and a maximum of 0.18 occurs at $\cos i_\mathrm{enc}\approx0.65$ ($i_\mathrm{enc}\approx 60^\circ$). The dispersion is large $\sim 0.1$ and some retrograde encounters barely capture. We emphasise that satellite capture is observed in 461 out of the 500 encounters, so it is not that most registered captures emerge in a few encounters while most encounters lead to no capture. Also, in the plot, these encounters are roughly evenly distributed in $\cos i_\mathrm{enc}$, a hint of an isotropic distribution of the encounters \citep[cf.][]{Li2020}.

\begin{figure*}
\centering
\includegraphics[width=0.85\hsize]{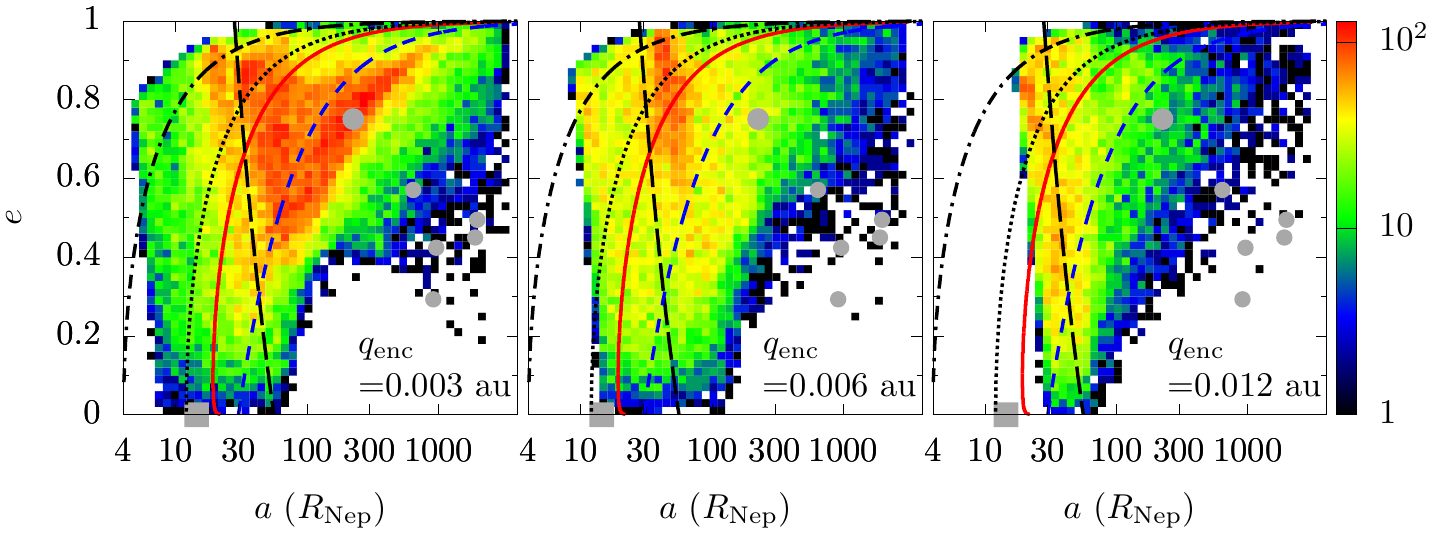}
\caption{Distribution of $(a,e)$ of the orbits of the captured moons from the inner moon region of IG18 at different encounter distances. The encounter distances are given in the bottom right corner of each panel. Warmer colours mean higher numbers, and grey dots are the observed satellite (cf. Figure \ref{fig-ae-moon-1-dist}). The red solid line indicates a tidal circularisation isochrone of 4 Gyr; the blue short-dashed line represents an equal pericentre distance $q=30R_\mathrm{Nep}$; the black long-dashed line delineates an equal apocentre distance of Nereid’s pericentre $Q=q_\mathrm{Ner}$; and the dash-dotted and dotted lines show tidal engulfment limit for prograde and retrograde orbits, respectively (cf. Figure \ref{fig-ae-illu}).}
 \label{fig-ae-moon-4-dist}
\end{figure*}

Then in Figure \ref{fig-ae-moon-4-dist}, we plot the 2D histogram of the distribution of the captured orbits in the $(a,e)$ plane for encounters with $q_\mathrm{enc}=$ 0.003, 0.006, and 0.012 au. When $q_\mathrm{enc}=0.003$ au (left), the orbits cover a wide range, from $a$ just above $4R_\mathrm{Nep}$ up to over $1000 R_\mathrm{Nep}$. $e$ is distributed over the entire range of $(0,1)$, with a preference for large values. Combined, the most favourable orbits form an asymmetric V-shaped region, with its bottom at around $(80 R_\mathrm{Nep},0.5)$, left tip at $(40 R_\mathrm{Nep},0.9),$ and right wing at $(500 R_\mathrm{Nep},0.9)$.

For the encounter distance of $q_\mathrm{enc}=0.006$ au (middle), the orbital distribution shrinks on both ends of $a$ (e.g. capture onto tight orbits with $a\lesssim 10 R_\mathrm{Nep}$ is now impossible). In addition, the V region contracts to a much smaller and less denser one around $a=50 R_\mathrm{Nep}$ and $e \gtrsim 0.7$. At more distant $q_\mathrm{enc}=0.012$ au (right), the region is confined more severely.

In the following two subsections, we use the reasoning in Section \ref{sec-class} to further classify the captured moons. See Figure \ref{fig-ae-illu} for our classification scheme.

\subsection{Survivable circularisable moons: Triton}

We start with the moons whose the orbits can be circularised by the   tides-only model. Those moons are in the red-shaded region in the left panel of Figure \ref{fig-ae-illu}. Their tidal circularisation time is shorter than the age of the Solar System and their orbits do not intersect that of Nereid.

The fraction of these moons, as shown in the Col. 4 of Table \ref{tab-cap-num-IG18}, accounts for roughly 1/5 of all captures when $q_\mathrm{enc}\lesssim 0.01$ au. The capture rate for these moons for each encounter as a function of $i_\mathrm{enc}$ is shown as the red dots  in the bottom panel of Figure \ref{fig-enc_inc_rate_sort} for $q_\mathrm{enc}=0.003$ au. The rate is $\lesssim 0.05$ in general analogous to that for all captures, and prograde encounters can more effectively capture.

As Figure \ref{fig-ae-moon-4-dist} shows, for the encounter distance $q_\mathrm{enc}=0.003$ au (left), part of the left tip of the high probability V-region can be circularised by tides only (to the left of both the red solid and the long-dashed lines). For $q_\mathrm{enc}=0.006$ au (middle) and $q_\mathrm{enc}=0.012$ au (right), significantly smaller fractions are circularisable.

When considering the further evolution after orbital circularisation, the close-in objects will be lost due to tidal engulfment by Neptune. In Figure \ref{fig-ae-moon-4-dist}, this means that those on the left side of the dash-dotted (for prograde orbits) and dotted lines (retrograde) cannot survive.

In Table \ref{tab-cap-num-IG18}, Cols. 5 (assuming prograde) and 6 (retrograde)   show the numbers of survivable circularisable moons in the tides-only model. During the closest encounters at $q_\mathrm{enc}=0.003$ au, such moons account for $\approx$$10\%$ of all captures. In other words, we have a  1\% chance of capturing a moon onto an orbit circularisable by tides only, without being engulfed by Neptune or damaging Nereid. During encounters of $q_\mathrm{enc}=0.006$ au, the chances decrease by $\lesssim 10\%$ and at $q_\mathrm{enc}=0.012$ au, the odds drop by another 60\%.

Then we check the situation where the tides+collisions model is functioning. Moons in the blue region in the right panel of Figure \ref{fig-ae-illu} can be circularised by this model. The fraction of circularisable moons through this model (the 8th column of Table \ref{tab-cap-num-IG18}) roughly triples or quadruples compared with those for tides-only model (the 4th column) and indeed up to 1/2 of all the captures are circularisable.

The blue short-dashed  line in Figure \ref{fig-ae-moon-4-dist} is the limit for circularisation by tides+collisions model. Compared to the tides-only model, significantly more moons on wider orbits become circularisable now, incorporating the entire left side of  the V-region for $q_\mathrm{enc}=0.003$ au as well as the high-density region for $q_\mathrm{enc}=0.006$ au.

We then remove those lost due to tidal engulfment by Neptune from the circularisable population and obtain the fraction of survivable moons as in the 9th and 10th columns of Table \ref{tab-cap-num-IG18}. The probability of obtaining a survivable circularisable moon increases by a factor of several compared to the tides-only model (the 5th and 6th columns). At $q_\mathrm{enc}=0.003$ au, the capture efficiency is $\sim4.4\%$ for survivable circularisable moons now.

We plot the capture efficiency of survivable circularisable moons in the two models as a function of their initial semimajor axes in Figure \ref{fig-a_cap_rate}. The red-slashed region is for tides only and the blue-slashed for collisions+tides; the upper (dash-dotted) and lower (dotted) bounds correspond to tidal engulfment for prograde and retrograde orbits, respectively. Both curves largely follow that for general capture (thick solid lines) if $a_\mathrm{ini}$ is small compared to $q_\mathrm{enc}$. However,  when $a_\mathrm{ini}\gtrsim q_\mathrm{enc}$, collisions are usually needed to circularise the orbit of a captured moon since these moons are often captured onto wide orbits, beyond the reach of tides or posing a threat to Nereid. 

Next, we comment on the inclination of these circularisable moons which, as discussed in Section \ref{sec-class}, is measured against the invariant plane. The distribution is shown in the middle panel of Figure \ref{fig-enc_inc_rate_sort} for each encounter as a function of its inclination $i_\mathrm{enc}$, and is represented by the median (red dots) and the 5th and 95th percentiles (grey horizontal line segments). Here we only show the result for moons circularisable by tides only and for encounter distance $q_\mathrm{enc}=0.003$ au. A major observation is that the inclination distribution covers a wide range, from prograde to retrograde configuration; thus it is not strange Triton happens to be retrograde. Additionally, it seems that when $\cos i_\mathrm{enc}\gtrsim0.6$, the median of $i_\mathrm{CM}$ is monotonically increasing from $\sim20^\circ$ to $\sim 100^\circ$ as $\cos i_\mathrm{enc}$ decreases (meaning that $i_\mathrm{enc}$ rises).

\begin{figure}
\centering
\includegraphics[width=\hsize]{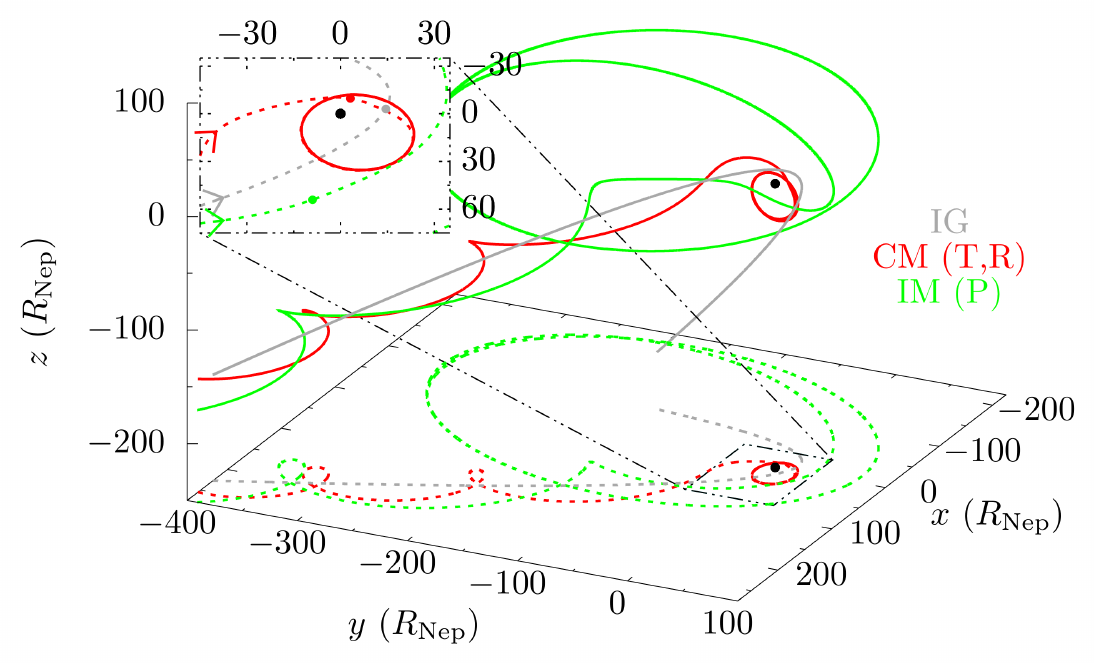}
\caption{Trajectories of two moons captured by Neptune and of the ice giant (IG) during the  encounter of the two planets. Neptune is the black dot at the origin. The grey line represents the path of the IG at $q_\mathrm{enc}=0.003$ au, red that of a retrograde circularisable (by tides only) moon [CM~(T,R)], and green that of a prograde irregular moon [IM~(P)]. Both moons are initially orbiting the IG, entering the plotted area from the bottom left, and then captured by Neptune. The projections onto the $x$-$y$ plane (dotted lines) are also shown with a zoomed-in view in the top left corner of the plot where the dots show the positions of the objects at the   closest approach of the two planets and the arrows the directions of motion.}
\label{fig-xyz}
\end{figure}

Finally, we show an example trajectory of a moon circularisable by tides during an encounter at $q_\mathrm{enc}=0.003$ au in Figure \ref{fig-xyz}. Here, Neptune is fixed at the origin (the black dot), and the moon's path is shown in red and that of the IG in grey. The projection of the trajectories onto the $x$-$y$ plane (the invariant plane) is shown with a zoomed-in view, where the circles mark the positions of the objects at the   closest approach of the two planets  in the top left corner. We have deliberately chosen a moon on a retrograde orbit. Upon capture, this circularisable moon happens to approach Neptune on a side different from that of the IG. Since the trajectory of this IG is prograde, this circularisable moon enters a retrograde orbit.

\subsection{Irregular moons: Nereid}

We define irregular moons as those uncircularisable in either of the two circularisation models and with $a>70R_\mathrm{Nep}$ (cf. Figure \ref{fig-ae-illu}). Along with the two models of orbital circularisation, we have two populations of irregular moons as listed in Cols. 7 (tides only) and 11 (collisions+tides)  of Table \ref{tab-cap-num-IG18}. During our closest encounters at $q_\mathrm{enc}=0.003$ au, $\sim4\%$ of the moons around IG18 are captured by Neptune onto irregular moon orbits. For more distant encounters, the fractions for irregular moons decrease and they drop faster than those for general capture.

The orbital distribution of these irregular moons can be seen in Figure \ref{fig-ae-moon-4-dist}, located on the right side of the red solid line (circularisation by tides only) and a vertical line $a=70 R_\mathrm{Nep}$ (the green-checked region in the left panel of Figure \ref{fig-ae-illu}) or to the right of the blue short-dashed line (by collisions + tides) and the line $a=70 R_\mathrm{Nep}$ (the green-checked region in the right panel of Figure \ref{fig-ae-illu}). Obviously, these orbits differ significantly from the observed irregular satellites (grey circles). Most of our irregulars have $a \sim 100 R_\mathrm{Nep}$, whereas the real ones are further out. Only a small fraction of our modelled moons end up with $a \gtrsim1000 R_\mathrm{Nep}$, but even   their eccentricities are too high to match the observations. This is the same for all encounter distances; they are  all unable to capture moons onto orbits with $a \gtrsim1000 R_\mathrm{Nep}$ and moderate eccentricities. So for the parameter ranges explored here, the capture of irregular satellites from a flying-by planet's inner moon region is not an effective candidate mechanism.

Nonetheless, Nereid (large grey circle in Figure \ref{fig-ae-moon-4-dist}) is   right in the middle of the V-shaped high-density region of the captures for $q_\mathrm{enc}=0.003$ au. Such encounters tend to create objects on $a \sim 100 R_\mathrm{Nep}$ and $e\gtrsim0.5$ orbits, the exact orbital feature of Nereid. At further encounter distances, the high-density region moves away from Nereid and when $q_\mathrm{enc}\gtrsim0.01$ au, Nereid-like orbits cannot be  created easily.

As before, we plot the capture rate as a function of the initial semimajor axis for the irregular moons (the green-checked regions in Figure \ref{fig-a_cap_rate}, the upper limit by the tides-only circularisation model and the lower by collisions+tides). As can be seen, unlike the capture rate for survivable circularisables, the capture rate for irregulars is monotonically increasing. Also, irregular moons can only be captured when $a_\mathrm{ini}\gtrsim a_\mathrm{enc}/2$, slightly stricter than for survivable circularisable moons.

Then the capture rate for the irregular moons in each encounter as a function of the encounter inclination $i_\mathrm{enc}$ is shown in the bottom panel of Figure \ref{fig-enc_inc_rate_sort} (green dots) for $q_\mathrm{enc}=0.003$ au. The same as those for all captured moons (black) and for circularisable moons (red), there is a preference for prograde encounters and the dispersion is large.

The inclination distribution of the irregulars captured in each encounter is presented in the top panel of Figure \ref{fig-enc_inc_rate_sort} as a function of $i_\mathrm{enc}$ for $q_\mathrm{enc}=0.003$ au. As for the circularisable moons (middle panel), here we have the median (green dots) and the 5th and the 95th percentiles (grey horizontal line segments). The median displays large dispersion, varying from encounter to encounter. The 5th and 95th percentiles also show significant scattering, but the ranges spanned by the two are wide.

The trajectory of a prograde irregular moon is shown in Figure \ref{fig-xyz} in green. Opposite to the circularisable moon (in red), this moon is approaching Neptune on the same side as the IG whose Neptune-centric path is prograde. Thus, it acquires a prograde orbit around Neptune when captured.

\section{Captured moons at Neptune from the inner moon region around IG10}\label{sec-res-IG10}
In the previous section, we have analysed in detail the capture efficiency and the characteristics of the captured orbits for encounters between Neptune and IG18. In this section we report the statistics of captures from the inner moon region of IG10; Table \ref{tab-sim-sum} lists the simulation parameters.

The capture rates for different types of moons at different encounter distances are summarised in Table \ref{tab-cap-num-IG10}. Compared to the case of IG18, loss and capture efficiencies increase by 20\% for our closest encounters penetrating to the inner moon region. Further out, i.e. beyond the outer edge of the initial moon distribution, Neptune can capture significantly more moons from IG10 than from IG18 but the numbers are always small. Average efficiencies for IG10 and IG18 differ only by a few tens of \%. This is not surprising, as the capture process does not strongly depend on the mass ratio between the new and the original host planets \citep[][and see also \citealt{Hills1989,Pfalzner2005,Jilkova2016}]{Hong2018}.

The orbital distribution of captured moons from IG10 is similar to those from IG18 (Section \ref{sec-res-IG18}) and our classification statistics are proportionately similar (Table \ref{tab-cap-num-IG10}). Up to several \% of the moons are captured as survivable circularisable moons or irregular moons, depending on the encounter distance and the orbital circularisation mechanism.

\begin{table*}[h]
\caption{As Table \ref{tab-cap-num-IG18} but for captures from the inner moon region around IG10.}
\begin{center}
\label{tab-cap-num-IG10}
\begin{tabular}{ccc|cccc|cccc}
\hline
\hline
 \multirow{ 2}{*}{$q_\mathrm{enc}$ (au)}&\multirow{ 2}{*}{loss (\%)}& \multirow{ 2}{*}{capture (\%)} & \multicolumn{4}{c|}{tides only (\%)} & \multicolumn{4}{c}{tides+collisions (\%)}\\
&&&CM&SCM (P)& SCM (R) &IM& CM&SCM (P)& SCM (R) &IM\\
\hline
0.003&62.92&11.81&2.68&2.24&0.88&5.91&6.97&6.55&4.64&4.39\\
0.006&42.70&8.13&1.51&1.35&0.65&2.75&5.71&5.46&4.26&1.83\\
0.012&14.10&5.16&0.38&0.37&0.20&2.04&3.08&3.00&2.55&1.32\\
0.024&0.29&0.23&0&0&0&0.1&0.08&0.07&0.07&0.08\\
\hline
\end{tabular}
\end{center}
\end{table*}
\section{Captured moons at Neptune from the outer moon region around IG18}\label{sec-primo-irre}
So far we have studied the moons captured by Neptune from the IG's inner moon region. Captures can be classified into two categories: circularisable moons, where the orbits shrink and circularise within the age of the Solar System, and irregular moons left with eccentric orbits. In this section, we show results from the outer moon region of IG18; Table \ref{tab-sim-sum} lists the simulation parameters.

The loss and capture rate at all eight IG--Neptune encounter distances are listed in Table \ref{tab-primo-irre}. During the closest encounters at $q_\mathrm{enc}=0.003$ au, nearly 80\% of the outer moons are lost from IG18 and 40\% are captured by Neptune. Up to $q_\mathrm{enc}=0.024$ au, the loss rate decreases by only 20\%, while that for capture by 70\%. Even so, nearly 10\% of the outer moons are acquired by Neptune. For more distant encounters, the two rates decline faster, reaching 10\% and 1\% at 0.1 au, respectively.

\setlength{\tabcolsep}{0.5em} 
\begin{table}[h]
\caption{Loss and capture rates for moons from the outer moon region around IG18 at different encounter distances. In Cols. 1 and 4 we list the encounter distance, in Cols. 2 and 5 the loss rates, and  in Cols.  3  and 6 the capture rates.}
\begin{center}
\label{tab-primo-irre}
\begin{tabular}{ccc|ccc}
\hline
\hline
$q_\mathrm{enc}$ (au)&loss (\%)&cap (\%)&$q_\mathrm{enc}$ (au)&loss (\%)&cap (\%)\\
\hline
0.003&86.4&36.5&0.048&41.66&2.50\\
0.006&85.3&27.3&0.096&13.49&0.64\\
0.012&78.0&19.5&0.192&1.31&0.17\\
0.024&67.0&9.28&0.384&0.08&0.00\\
\hline
\end{tabular}
\end{center}
\end{table}

\begin{figure}
\centering
\includegraphics[width=\hsize]{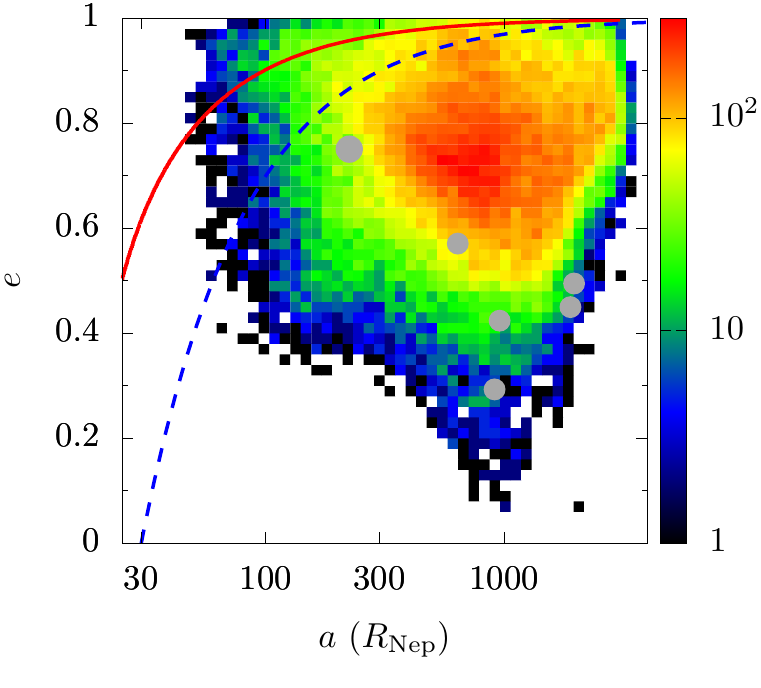}
\caption{Distribution of $(a,e)$ of the orbits of the captured moons from the outer moon region of IG18 at the encounter distance of $q_\mathrm{enc}=0.003$ au. Warmer colours mean higher numbers and grey dots are the observed satellite; cf. Figure \ref{fig-ae-moon-1-dist}. The red solid line marks a tidal circularisation isochrone of 4 Gyr; the blue (short-)dashed line represents an equal pericentre distance $q=30R_\mathrm{Nep}$; cf. Figure \ref{fig-ae-illu}.}
\label{fig-ae-primo-irre}
\end{figure}

Figure \ref{fig-ae-primo-irre} shows the distribution of capture orbits for encounters with $q_\mathrm{enc}=0.003$ au overplotted with the observed population. Compared to those captured from the inner region (Figure \ref{fig-ae-moon-4-dist}) these moons typically have wider orbits, mostly $\in(300,2000)R_\mathrm{Nep}$, much the same as the observed irregulars, but wider than that of Nereid. Nonetheless, their orbits are highly eccentric, predominantly $e\gtrsim 0.5$, and are  inconsistent with observations. For  encounters at greater distances, the inner boundary of $a$ increases, but $e$ is still too high compared to observations. The vast majority of captures are classified as irregular moons in both circularisation models, as per Section \ref{sec-class}, and circularisable moons cannot be captured from the outer moon region.

Hence, we conclude that while the capture efficiency from the outer moon region at the IG is higher than from the inner region, the captured orbits do not resemble the observed satellites. Even if moons are captured by Neptune from this region, they must either have been lost or their orbits have evolved significantly.
\section{Discussion}\label{sec-dis}
\subsection{Efficiency estimate}\label{sec-eff}
We concentrate on the capture of moons by Neptune from the inner region of the IG since the resulting orbits may resemble the observed moons Triton and Nereid while captures from the outer moon region may not.

Whether an  IG--Neptune deep close encounter has really occurred, which is  required for Neptune to capture from the inner moon region of the IG,  is unknown. Based on self-consistent full Nice scenario simulations, a few papers have looked into the detailed encounter history between an IG and Jupiter \citep{Nesvorny2014} and between an IG and Saturn \citep{Nesvorny2014a}. In these studies, 13 encounter sequences (10 at Saturn and 3 at Jupiter) from different runs were presented. Here an encounter sequence means all close encounters experienced by a planet (e.g. Saturn) in one run of Nice scenario simulation. The criterion for the selection of these runs was that the correct Solar System configuration was reproduced. So the encounters were probably not biased with regard to themselves in the sense that they were chosen not according to the encounter geometry. These authors showed the encounter distance of the closest one for each of the 13 sequences. Three are closer than 0.01 au ( at 0.003 au,  0.007 au, and  $\approx0.01$ au); in another three sequences, the deepest ones reach $\lesssim0.02$ au. To our knowledge, no works on statistics of encounters between an IG and Neptune have been published. So we assume that IG--Neptune encounters are similar to those involving IG--Saturn and IG--Jupiter. Thus, roughly speaking, an IG--Neptune encounter within 0.02 au has an occurrence rate of $50\%$ (6/13) and that inside 0.01 au of $25\%$ (3/13).

Satellite formation at an ice giant is not well understood. Uranus has a well-defined regular moon system, with four major satellites located between 8 and 23 Uranian radii, each about half the size of Triton in radius. There is evidence that moons further away than Oberon (the outermost of the four) would have been removed during the Nice scenario \citep{Deienno2011}. So Uranus might have a more extended satellite system in the past. Recently, \citet{Szulagyi2018} showed that Triton-sized moons could actually form around ice giants and possibly on a wide range of semimajor axes. Thus, if IG18 has four moons at random locations between $5R_\mathrm{Nep}$ and $60R_\mathrm{Nep}$, the chance for Neptune to capture one is $4\times 9.5\%=38\%$ during an encounter at $q_\mathrm{enc}=0.003$ au, $25\%$ at 0.006 au, $14\%$ at 0.012 au, and $0.2\%$ at 0.024 au (Table \ref{tab-cap-num-IG18}). The mean rate for the four distances is 19\%.

When multiplying these rates with that of the encounter actually happening (50\%) the combined probability is 10\% for Neptune to capture one  of the four moons in the inner moon region of IG18. However, if much like the current Uranian system, all of IG18's four moons form inside $30R_\mathrm{Nep}$, this possibility drops to 3\%(cf. Figure \ref{fig-a_cap_rate}). In a similar way, the probability increases if IG18 has more moons on more distant orbits.

Applying the same analysis to the inner moon region of IG10, we estimate that, if IG10 initially has four moons $\in(5R_\mathrm{Nep},60R_\mathrm{Nep})$, the chance for Neptune to capture one is 13\%;  if within $\in(5R_\mathrm{Nep},30R_\mathrm{Nep})$ the chance is 4\% (cf. Table \ref{tab-cap-num-IG10}).

However, whether a captured moon at Neptune is later circularised by tides-only or by collisions+tides and survives tidal engulfment, becoming a survivable circularisable moon on a close-in circular Triton-like orbit, or remains on a wide and eccentric orbit turning into an irregular moon like Nereid is highly model and/or parameter dependent. Hence, we choose not to overinterpret our results. Nonetheless, the discussion in Sections \ref{sec-class}, \ref{sec-res-IG18}, and \ref{sec-res-IG10} clearly shows that  circularisable and irregular moons can be captured in the same encounter, and  so can prograde and retrograde moons. We note again that our model cannot create captures on wide but only moderately eccentric trajectories: orbital features of the extant irregular satellites cannot be reproduced, and the observed irregular satellites have   probably been populated via other mechanisms \citep{Nesvorny2007}, but the two moons with unique orbital features can be explained by our model. Figure \ref{fig-xyz} shows how a retrograde circularisable moon (e.g. Triton) and a prograde irregular moon (e.g. Nereid) are simultaneously created.

The capture efficiency of Neptune from the outer moon region of the IG is much higher and capture is possible during much more distant encounters. However, the resulting orbits do not match any of the observed moons. We are also  not aware of a statistical study on these distant encounters within the Nice scenario in the literature. So we omit estimating their efficiency.

\subsection{Comparison with other models}

Several models have been proposed for either capture \citep{Agnor2006,Goldreich1989,McKinnon1995,Pollack1979} or in situ formation \citep{Harrington1979,Li2020} for Triton. A widely accepted capture model is that of exchange capture \citep[][and see Section \ref{sec-intro}]{Agnor2006}. This model features a high capture efficiency of 50\% if hundreds of pairs of very wide Triton-sized binaries are present in the reservoir, the primordial planetesimal disc \citep[][but in another simulation, \citealt{Vokrouhlicky2008} reported an efficiency of 2\%]{Nogueira2011}. Or Triton may be Neptune's primordial regular moon and is reverted onto a retrograde orbit by an IG encounter \citep[][and cf. \citealt{Harrington1979}]{Li2020}. Though  at a much lower rate of $\sim10^{-5}$ and requiring specific and possibly not-so-realistic initial conditions, this model can explain Triton's full orbital evolution until circularisation in a self-contained and self-consistent way.

Regarding Nereid, capture from the planetesimal disc \citep{Nesvorny2007,Nesvorny2014} is among the best candidate mechanisms at a likelihood of $\sim$1\% \citep{Li2020}. \citet{Li2020} showed that Nereid-like wide and eccentric orbits can be naturally created from initial close-in circular ones during an IG encounter \citep[see also][]{Harrington1979} and the efficiency is $\sim$1\%. The exchange capture model, as advocated for Triton \citep{Agnor2006}, may have its efficiency peaked at Nereid's size \citep[][and see also \citealt{Philpott2010}]{Vokrouhlicky2008}, but dedicated simulations for its capture at Neptune are missing.

These different models, to some extent, all looked at different aspects of Triton and/or Nereid with various assumptions and parameters or actually not designed to explain the two moons in the first place. For example, the mechanism of \citet{Nesvorny2007} was proposed not to account for the specifics of Nereid, but to reproduce the overall population of the observed irregular satellites and was highly successful. To make an objective comparison is beyond the scope of this work. Here we only point out a few implications of our model.

Our scenario does not depend on a preexisting Neptunian satellite system. As Figure \ref{fig-ae-moon-4-dist} and Table \ref{tab-cap-num-IG18} show, there is a fair chance ($\sim$$10\%$  of all captures) to capture Triton directly onto a survivable circularisable orbit inside that of Nereid. So Nereid does not need to be specifically protected and no collisions between Triton and the original moons are required.

In principle, Neptune may have captured satellites from other planets (e.g. Uranus) if the two planets encounter each other at close distances. However, such encounters, in addition to Neptune capturing satellites from Uranus, often lead to the satellites of Uranus  being destabilised to a large extent, and the survivors being highly excited \citep[Table \ref{tab-cap-num-IG18}, and see][]{Nesvorny2014a,Deienno2014,Li2020}. As mentioned in Section \ref{sec-intro}, this violates the fact that all major moons of Uranus (and other giant planets) are dynamically cold. In contrast, we have no constraints from the IG's side simply because it has been ejected later and not observed. Also, in the jumping Jupiter variant of the Nice scenario, which  better reproduces the Solar System observations, the surviving four giant planets all seem to   have mild orbital evolution during the instability period \citep[e.g.][]{Nesvorny2018}. This suggests that the requirement of high eccentricities for these planets to cross each other's orbit is likely not met, so close encounters between them are implausible. Hence, our proposed Neptune's capture of satellites from another planet is only consistent with the jumping Jupiter with three IGs.

\subsection{Model limitation}\label{sec-cav}

Our model is linked to a period of early Solar System instability \citep{Nesvorny2012} where the IG, possibly due to encountering mean motion resonances with Jupiter and Saturn, is orbitally excited, physically encounters those planets and is then flung outwards \citep{Tsiganis2005} towards Neptune. It follows then that our model depends on the IG’s satellites surviving these encounters \citep[e.g.][]{Deienno2011,Gomes2012}. One constraint discussed earlier is that both Jupiter and Saturn have radially extended regular satellite systems, and the encounters cannot be too close. For Jupiter, planetary encounters deeper than 0.02 au are ruled out so as to avoid overexciting the Galilean moons, while those between 0.03 and 0.05 au only cause small perturbations \citep{Deienno2014}. In a similar vein, encounters within 0.02 au of Saturn must be avoided \citep[though 0.05 au is a safer limit;][]{Nesvorny2014a}.

We have performed additional simulations to examine how effective these encounters are in destabilising or outright ejecting the IG satellites. Taking Jupiter and IG18 as a case in point, we generate 100 encounters at $d_\mathrm{enc}=0.024$ as in Section \ref{sec-sim}, and integrate a total of 1100 test satellites of the IG between $5R_\mathrm{Nep}$ and $60R_\mathrm{Nep}$ through the encounters. The overall moon survival rate is 83\%, yet the orbits are excited significantly; the median eccentricity of all surviving test moons is 0.2 and a fraction of the surviving moons acquire wider orbits, which are unlikely to subsequently produce the observed satellites at Neptune (Section \ref{sec-primo-irre}). Furthermore, orbital excitation in a multi-satellite system would lead to disruptive mutual collisions that deplete the source population. However, we also find that test moons orbiting the IG inside $30R_\mathrm{Nep}$ are effectively stable, with a survival rate of 99\% and median eccentricity of 0.04. Therefore, we suspect that the inner part of the inner moon region is immune to the disturbing influence of the gas giants. In addition, even if primordial moons at small planetocentric distances may be collisionally disrupted, the resulting debris disc would damp to the local Laplace plane and new moons would reaccrete quickly on timescales of hundreds of years \citep{Banfield1992,Rufu2017}. If we restrict ourselves to moons inside $30R_\mathrm{Nep}$, we find that the model efficiency drops by a factor of three, but is still several per cent in absolute terms.

Having established the availability of suitable source objects around the IG, we now consider whether the IG--Neptune encounter is consistent with the existence of Neptunian moons other than Triton and Nereid. \citet{Li2020} studied this problem and concluded that even for encounters at 0.003 au between IG18 and Neptune, the orbit of Proteus, the inner neighbour of Triton, is only slightly excited, while the outer irregular satellites may be captured later in subsequent encounters \citep{Nesvorny2007}. In conclusion, the current satellite configuration at Neptune is compatible with our model.

Finally, can our captured moons survive? Neptune has probably experienced more than one encounter with the IG \citep{Nesvorny2007,Nesvorny2014}. The one leading to the transfer of the moons from the IG to Neptune, as studied in this work, may not be the last. So can a captured moon withstand later planetary encounters? While \citet{Nesvorny2007} found that earlier-captured irregular satellites could not survive such events, \citet{Li2017a} showed that the Jovian irregular satellite Himalia, at about 0.2 Hill radii from its host, could probably survive $\sim$100 IG encounters. Since Nereid is deep within 0.05 Neptunian Hill radii \citep[cf. ][]{Li2020}, we suspect that it probably does survive, and so does Triton which is much deeper within Neptune's potential well. On the other hand, the moons captured by Neptune from the outer moon region of the IG, because of their wide captured orbits, are probably lost in these later encounters. This could be a reason why we do not see these wide and highly eccentric objects now (Figure \ref{fig-ae-primo-irre}).

\subsection{Perspective of moon capture in exoplanetary systems}
Is the mechanism studied in this work applicable to extrasolar systems? As analysed before, the key elements in our model are close encounters between planets and preexisting satellite systems.

Thousands of exoplanets have been confirmed, showing diverse orbital features distinct from our own Solar System. Many are characterised by high eccentricities \citep{Udry2007}, suggestive of ubiquitous strong planetary scattering \citep{Rasio1996} during which  close encounters must be not uncommon \citep{Marzari2002}.

Despite years of effort \citep[e.g.][]{Kipping2012}, no convincing detections of exomoons have been reported yet \citep{Teachey2018a,Heller2018}. On the other hand, satellite formation may be a spontaneous consequence of the formation of the planet \citep[see a recent review by][]{Barr2017}. So, detections of exomoons may be just around the corner \citep{Teachey2018}.

A number of works have explored the stability of exomoons, either in a general three-body problem setup \citep{Domingos2006}, under tidal effects \citep{Barnes2002}, during the planets' migration \citep{Namouni2010}, or in the context of interplanetary scattering \citep{Gong2013}. \citet{Hong2018} recently showed that during the scattering, a small fraction of the moons can be transferred from one planet to another.

Further to these studies, our results indicate the following:
\begin{itemize}
\item A moon can jump from its original host planet to another, usually onto highly eccentric orbits.
\item Due to post-capture evolution, some of the moons can be circularised onto close-in orbits, while others may remain on wide and eccentric trajectories.
\item Moons on highly inclined orbits are naturally created, and the occurrence rates for prograde and retrograde captures are comparable.
\end{itemize}

\section{Conclusion}\label{sec-con}
During the early Solar System evolution, there is an instability period where planets fly by each other at close distances. In this paper we have explored,  during an encounter between an ice giant (IG) and Neptune, whether the IG's moons can be captured onto Neptune-bound orbits. We have examined the efficiency of moon capture at different encounter distances, finding that up to a few tens of per cent of the IG's moons can be transferred to Neptune during a single deep encounter. The overall capture probability, accounting for different encounter distances and the corresponding occurrence rates for the encounters, is estimated to be in the range 3\%-13\%..

Most of the captured satellites acquire elongated (and wide) orbits. Up to more than half of the captures can be circularised either by tides only or by collisions+tides, ending up on circular and close-in orbits, like that of Triton. Those not circularisable may stay on wide and eccentric orbits, becoming Nereid analogues. Both populations have wide distributions in orbital inclinations and the prograde and retrograde orbits have comparable occurrence rates. Thus,  if captured via this mechanism, Triton naturally has a 50\% chance to be counter-rotating.

We suggest that the above scenario may occur in an exoplanetary system if primordial exomoons are formed and the system becomes unstable.

\begin{acknowledgements}
The authors are grateful to Rogerio Deienno, the referee, for constructive comments and suggestions. D.L. acknowledges financial support from Knut and Alice Wallenberg Foundation through two grants (2014.0017, PI: Melvyn B. Davies and 2012.0150, PI: Anders Johansen). A.J. thanks the Swedish Research Council (grant 2018-04867), the Knut and Alice Wallenberg Foundation (grants 2012.0150, 2014.0017) and the European Research Council (ERC Consolidator Grant 724687-PLANETESYS) for research support. 
\end{acknowledgements}

\begin{appendix}
\section{Tidal evolution prescription}\label{sec-tid}
In isolation, two celestial bodies orbit their centre of mass following Keplerian motion. However, they both have finite volumes and induce tidal distortions in each other. These tidal deformations perturb the two-body orbit in the long term, usually damping eccentricity and causing orbital drift. Here we follow an equilibrium tidal model \citep{Hut1981} to track the orbital evolution of the Neptune-moon system post-capture. In this model, the time lag between the phase of the tidal deformation and the Keplerian motion is constant.

\subsection{Damping of eccentricity}

Before orbital circularisation, the distortion in the planet caused by the moon plays a minor role and is omitted. Following \cite{Correia2009}, due to the deformation in the moon, the time evolution of orbital eccentricity $e$ can be approximated as
\begin{equation}
\label{eq-e}
\dot e={9K_\mathrm{Tri} e \over m_\mathrm{Tri}a^2}\left[{11\over18} f_4(e)\cos\theta_\mathrm{Tri} {\omega_\mathrm{Tri} \over n}-f_5(e) \right]
,\end{equation}
where $m_\mathrm{Tri}$ is the moon's mass, $a$ the semimajor axis, $\theta_\mathrm{Tri}$ the moon's obliquity, $\omega_\mathrm{Tri}$ the moon's spin rate, $n$ the mean motion, and $f_4(e)$ and $f_5(e)$ functions of $e$. The coefficient $K_\mathrm{Tri}$ is defined as $K_\mathrm{Tri}=3k_\mathrm{2,Tri}Gm^2_\mathrm{Nep}R^5_\mathrm{Tri}\Delta t_\mathrm{Tri}/a^6$ in which $k_\mathrm{2,Tri}$ is the moon's Love number, $G$ the gravitational constant, $m_\mathrm{Nep}$ Neptune's mass, $R_\mathrm{Tri}$ the moon's radius, and $\Delta t_\mathrm{Tri}$ the time lag, equal to  $\Delta t_\mathrm{Tri}=1/\omega_\mathrm{Tri}Q_\mathrm{Tri}$ ($Q_\mathrm{Tri}$ is the moon's tidal parameter). Along with $e$, $a$, $\theta_\mathrm{Tri}$, and $\omega_\mathrm{Tri}$ are also evolving with time and we omit the equations of motion for them.

The synchronisation of the moon's spin with its orbital motion is much faster than the evolution of $a$ and $e$ in that $\omega_\mathrm{Tri}$ evolves to $n$ and $\theta_\mathrm{Tri}$ approaches $0$ much more quickly than the evolution of $n$. Hence, we can substitute the two and Equation \eqref{eq-e} only depends on  $a$ and $e$.

The total angular momentum of the system, i.e. the sum of the orbital and the moon's spin angular momenta, is rigorously conserved; however, that carried by the moon's rotation is negligible due to its small size so the orbital angular momentum itself is quasi-conserved. Hence, the normalised orbital angular momentum
\begin{equation}
\label{eq-am}
\Lambda=\sqrt{a(1-e^2)}
\end{equation}
is a constant. Solving Equation \eqref{eq-am} for $a$ and substituting it into Equation \eqref{eq-e} (noting $\omega_\mathrm{Tri}=n$ and $\theta_\mathrm{Tri}=0$), we have
\begin{equation}
\label{eq-e-2}
\dot e=-C_\mathrm{Tri} F_\mathrm{Tri} (e)
,\end{equation}
where
\begin{equation}
\label{eq-C-e}
C_\mathrm{Tri}={27 G  k_\mathrm{2,T} m^2_\mathrm{Nep}R^5_\mathrm{Tri} \Delta t_\mathrm{Tri}\over \Lambda^{16} m_\mathrm{Tri}}
\end{equation}
and $F_\mathrm{Tri}(e)$ is a function of $e$.

Thus, in essence, the temporal evolution of $e$ is governed by $F_\mathrm{Tri}(e)$ on timescales modulated by the constant-coefficient $C_\mathrm{Tri}$ which is determined by $\Lambda$ and model assumptions. In other words, $e$ of any moon, under this model, follows exactly the same evolutionary path that is only stretched in time according to the moon's $C_\mathrm{Tri}$. This means that we only need to perform one numerical integration of the equation of motion \eqref{eq-e-2}, leaving $C_\mathrm{Tri}$ aside. And by applying the specific $C_\mathrm{Tri}$ for each moon, we obtain their evolution in $e$ and hence the orbital circularisation time.

The physical and tidal parameters of a captured moon are unknown. For simplicity, we adopt Triton's values \citep{Goldreich1989,Correia2009,Nogueira2011}. Specifically, we assume the moon is rocky, with $Q_\mathrm{Tri}=100$. The moon may be (semi-)molten owing to the energy dissipated in it \citep{McKinnon1984,Goldreich1989}, which will accelerate $e$-damping by reducing $Q_\mathrm{Tri}$, potentially speeding up the tidal evolution by a factor of 10. We do not take this into account because many uncertainties are involved. For example, Kozai--Lidov cycles \citep{Kozai1962,Lidov1962} may induce large amplitude oscillations in $e$. This makes $\Lambda$ vary periodically, and may indeed increase circularisation time \citep{Cuk2005} by an order of magnitude.

Finally, we have performed a check against simulations by \citet{Li2020} where both distortions in the moon and Neptune were considered and the agreement is excellent.

\subsection{Orbital drift after circularisation}

After orbital circularisation, the tidal deformation in Neptune dominates and the moon may migrate inwards or outwards due to the exchange of orbital angular momentum with Neptune's spin angular momentum. If $a$ is small, the drift may be fast enough that the moon can be engulfed by Neptune within the age of Solar System (4 Gyr). We need to exclude these circularisable but not survivable moons when counting survivable CMs.

The equation of motion for $a$, governed by the tidal deformation in Neptune (assuming zero $e$), is \citep{Correia2009}
\begin{equation}
\label{eq-a}
\dot a=-{2K_\mathrm{Nep} \over m_\mathrm{Tri}a}\left[1-\cos \theta_\mathrm{Nep} {\omega_\mathrm{Nep} \over n} \right].
\end{equation}
Here $K_\mathrm{Nep}=3k_\mathrm{2,Nep}Gm^2_\mathrm{Tri}R^5_\mathrm{Nep}\Delta t_\mathrm{Nep}/a^6$, with $k_\mathrm{2,Nep}$   Neptune's Love number; $R_\mathrm{Nep}$ is the Neptunian radius;  $\Delta t_\mathrm{Nep}$ is the time lag expressed as $\Delta t_\mathrm{Nep}=1/\omega_\mathrm{Nep}Q_\mathrm{Nep}$, with  $Q_\mathrm{Nep}$   Neptune's tidal parameter; $\theta_\mathrm{Nep}$ is the inclination of the moon's orbit with respect to Neptune's equator; and $\omega_\mathrm{Nep}$ is Neptune's spin rate. Alongside $a$, $\theta_\mathrm{Nep}$, and $\omega_\mathrm{Nep}$ are also time-variable. 

As before, the sum of the orbital angular momentum and Neptune's spin angular momentum,
\begin{equation}
L_\mathrm{tot}={m_\mathrm{Tri}m_\mathrm{Nep}\over\sqrt{m_\mathrm{Tri}+m_\mathrm{Nep}}}\sqrt{Ga}+\omega_\mathrm{Nep} K m_\mathrm{Nep}R^2_\mathrm{Nep},
\end{equation}
is conserved. Here $K$ is a constant related to Neptune's rotational inertia and is of order unity \citep{Hubbard1991}. For a Triton-mass moon \cite[$0.2\%$ of that of Neptune;][]{Murray1999} it can be shown that $L_\mathrm{tot}$ is dominated by the rotation of Neptune. So in our application we assume that $\omega_\mathrm{Nep}$ is constant and adopt Neptune's current value.

From Equation \eqref{eq-a} it is evident that if $\theta_\mathrm{Nep}>90^\circ$ (i.e. retrograde as viewed with respect to Neptune's equator), the moon is always drifting inwards. Otherwise, the direction of the drift of $a$ may depend on the ratio $\omega_\mathrm{Nep}/n$. In our simulations $\theta_\mathrm{Nep}$ is not well defined as the direction of the spin axis of Neptune is not known.

For simplicity, we consider only two cases.  In the first, all moons are assumed to have $\theta_\mathrm{Nep}=0^\circ$. In this case, a moon with $a_\mathrm{cir}$ (semimajor axis upon circularisation) smaller than that of the synchronous orbit ($\approx 3.4 R_\mathrm{Nep}$) is lost to Neptune. However,  we strengthen the requirement to $4 R_\mathrm{Nep}$ to account for a more probable situation where there is a small inclination (and thus a faster drift speed).  In the second case all moons are assumed to have $\theta_\mathrm{Nep}=180^\circ$. Now those inside $\sim 14 R_\mathrm{Nep}$ are lost\footnote{Triton may be absorbed by Neptune in the next few Gyr \citep{Chyba1989}.}. Because most retrograde moos do not have an inclination of $180^\circ$, their inward drift should be slower. And during this drift, $\theta_\mathrm{Nep}$ will actually slowly decrease \citep{Chyba1989}, which decelerates the drift. Furthermore, the initial tidal circularisation phase may take a significant fraction of the age of the Solar System, so equating this drift time to 4 Gyr probably leads to an overestimate of tidal engulfment. Thus, we decide to set the criterion to $a_\mathrm{cir}=12 R_\mathrm{Nep}$. Above we  consider two extreme cases and the real situation should be in between; these two represent the upper and lower dispersion of tidal drift loss.

If tides are the only orbital circularisation factor, the orbital angular momentum is quasi-conserved. Things may be more complicated when collisions are involved. However, we note that a requirement for the collisions+tides model to work is that the captured moon itself is not collisionally disrupted, meaning that the moons hitting Triton cannot be too massive \citep{Rufu2017}. In addition, as shown in \citet{Li2020}, collisions with moons of a total mass of only a few per cent of Triton's are sufficient for its orbital decoupling from Nereid. Hence, the orbital angular momentum of Triton (or a CM) is quasi-conserved during collisional evolution as well. Therefore, here we  simply assume any CM's orbital angular momentum \eqref{eq-am} is conserved whether it is circularised through  tides only or through collisions+tides, and we use this relation to calculate $a_\mathrm{cir}$.


\end{appendix}

%

\begin{thebibliography}{75}
\expandafter\ifx\csname natexlab\endcsname\relax\def\natexlab#1{#1}\fi

\bibitem[{Agnor \& Hamilton(2006)}]{Agnor2006}
Agnor, C.~B. \& Hamilton, D.~P. 2006, Nature, 441, 192

\bibitem[{Banfield \& Murray(1992)}]{Banfield1992}
Banfield, D. \& Murray, N. 1992, Icarus, 99, 390

\bibitem[{Barnes \& O'Brien(2002)}]{Barnes2002}
Barnes, J.~W. \& O'Brien, D.~P. 2002, The Astrophysical Journal, 575, 8

\bibitem[{Barr(2017)}]{Barr2017}
Barr, A.~C. 2017, Astronomical Review, 12, 24

\bibitem[{Batygin {et~al.}(2012)Batygin, Brown, \& Betts}]{Batygin2012a}
Batygin, K., Brown, M.~E., \& Betts, H. 2012, Astrophysical Journal Letters,
  744, L3

\bibitem[{Beaug{\'{e}} {et~al.}(2002)Beaug{\'{e}}, Roig, \&
  Nesvorn{\'{y}}}]{Beauge2002}
Beaug{\'{e}}, C., Roig, F., \& Nesvorn{\'{y}}, D. 2002, Icarus, 498, 483

\bibitem[{Bou{\'{e}} \& Laskar(2010)}]{Boue2010}
Bou{\'{e}}, G. \& Laskar, J. 2010, Astrophysical Journal Letters, 712, L44

\bibitem[{Brasser {et~al.}(2009)Brasser, Morbidelli, Gomes, Tsiganis, \&
  Levison}]{Brasser2009}
Brasser, R., Morbidelli, A., Gomes, R., Tsiganis, K., \& Levison, H.~F. 2009,
  Astronomy and Astrophysics, 507, 1053

\bibitem[{Burns(1986)}]{Burns1986a}
Burns, J.~a. 1986, in Satellites, ed. J.~A. Burns \& M.~S. Matthews (Tucson:
  University of Arizona Press), 117--158

\bibitem[{Canup \& Ward(2002)}]{Canup2002}
Canup, R.~M. \& Ward, W.~R. 2002, The Astronomical Journal, 124, 3404

\bibitem[{Chambers(1999)}]{Chambers1999}
Chambers, J.~E. 1999, Monthly Notices of the Royal Astronomical Society, 304,
  793

\bibitem[{Chyba {et~al.}(1989)Chyba, Jankowski, \& Nicholson}]{Chyba1989}
Chyba, C.~F., Jankowski, D.~G., \& Nicholson, P.~D. 1989, Astronomy and
  Astrophysics, 219, L23

\bibitem[{Cloutier {et~al.}(2015)Cloutier, Tamayo, \& Valencia}]{Cloutier2015}
Cloutier, R., Tamayo, D., \& Valencia, D. 2015, The Astrophysical Journal, 813,
  8

\bibitem[{Correia(2009)}]{Correia2009}
Correia, A. C.~M. 2009, The Astrophysical Journal, 704, L1

\bibitem[{{\'{C}}uk \& Gladman(2005)}]{Cuk2005}
{\'{C}}uk, M. \& Gladman, B.~J. 2005, The Astrophysical Journal, 626, L113

\bibitem[{Dawson \& Murray-Clay(2012)}]{Dawson2012}
Dawson, R.~I. \& Murray-Clay, R. 2012, Astrophysical Journal, 750, 43

\bibitem[{Deienno {et~al.}(2017)Deienno, Morbidelli, Gomes, \&
  Nesvorn{\'{y}}}]{Deienno2017}
Deienno, R., Morbidelli, A., Gomes, R.~S., \& Nesvorn{\'{y}}, D. 2017, The
  Astronomical Journal, 153, 153

\bibitem[{Deienno {et~al.}(2014)Deienno, Nesvorn{\'{y}}, Vokrouhlick{\'{y}}, \&
  Yokoyama}]{Deienno2014}
Deienno, R., Nesvorn{\'{y}}, D., Vokrouhlick{\'{y}}, D., \& Yokoyama, T. 2014,
  The Astronomical Journal, 148, 25

\bibitem[{Deienno {et~al.}(2011)Deienno, Yokoyama, Nogueira, Callegari, \&
  Santos}]{Deienno2011}
Deienno, R., Yokoyama, T., Nogueira, E.~C., Callegari, N., \& Santos, M.~T.
  2011, Astronomy {\&} Astrophysics, 536, A57

\bibitem[{Domingos {et~al.}(2006)Domingos, Winter, \& Yokoyama}]{Domingos2006}
Domingos, R.~C., Winter, O.~C., \& Yokoyama, T. 2006, Monthly Notices of the
  Royal Astronomical Society, 373, 1227

\bibitem[{Goldreich {et~al.}(1989)Goldreich, Murray, Longaretti, \&
  Banfield}]{Goldreich1989}
Goldreich, P., Murray, N., Longaretti, P.~Y., \& Banfield, D. 1989, Science,
  245, 500

\bibitem[{Gomes {et~al.}(2018)Gomes, Nesvorn{\'{y}}, Morbidelli, Deienno, \&
  Nogueira}]{Gomes2018}
Gomes, R., Nesvorn{\'{y}}, D., Morbidelli, A., Deienno, R., \& Nogueira, E.
  2018, Icarus, 306, 319

\bibitem[{Gomes {et~al.}(2012)Gomes, Nogueira, \& Brasser}]{Gomes2012}
Gomes, R.~S., Nogueira, E.~C., \& Brasser, R. 2012, in DPS, 415.02

\bibitem[{Gong {et~al.}(2013)Gong, Zhou, Xie, \& Wu}]{Gong2013}
Gong, Y.-X., Zhou, J.-L., Xie, J.-W., \& Wu, X.-M. 2013, The Astrophysical
  Journal, 769, L14

\bibitem[{Harrington \& {Van Flandern}(1979)}]{Harrington1979}
Harrington, R.~S. \& {Van Flandern}, T.~C. 1979, Icarus, 39, 131

\bibitem[{Heggie(1975)}]{Heggie1975}
Heggie, D.~C. 1975, Monthly Notices of the Royal Astronomical Society, 173, 729

\bibitem[{Heller(2018)}]{Heller2018}
Heller, R. 2018, Astronomy {\&} Astrophysics, 610, A39

\bibitem[{Hills \& Dissly(1989)}]{Hills1989}
Hills, J.~G. \& Dissly, R.~W. 1989, The Astronomical Journal, 98, 1069

\bibitem[{Hong {et~al.}(2018)Hong, Raymond, Nicholson, \& Lunine}]{Hong2018}
Hong, Y.-C., Raymond, S.~N., Nicholson, P.~D., \& Lunine, J.~I. 2018, The
  Astrophysical Journal, 852, 85

\bibitem[{Hubbard {et~al.}(1991)Hubbard, Nellis, Mitchell, Holmes, McCandless,
  \& Limaye}]{Hubbard1991}
Hubbard, W.~B., Nellis, W.~J., Mitchell, A.~C., {et~al.} 1991, Science, 253,
  648

\bibitem[{Hut(1981)}]{Hut1981}
Hut, P. 1981, Astronomy and Astrophysics, 99, 126

\bibitem[{Hut \& Bahcall(1983)}]{Hut1983}
Hut, P. \& Bahcall, J.~N. 1983, The Astrophysical Journal, 268, 319

\bibitem[{J{\'{i}}lkov{\'{a}} {et~al.}(2016)J{\'{i}}lkov{\'{a}}, Hamers,
  Hammer, \& Zwart}]{Jilkova2016}
J{\'{i}}lkov{\'{a}}, L., Hamers, A.~S., Hammer, M., \& Zwart, S.~P. 2016,
  Monthly Notices of the Royal Astronomical Society, 457, 4218

\bibitem[{Kipping {et~al.}(2012)Kipping, Bakos, Buchhave, Nesvorn{\'{y}}, \&
  Schmitt}]{Kipping2012}
Kipping, D.~M., Bakos, G.~{\'{A}}., Buchhave, L., Nesvorn{\'{y}}, D., \&
  Schmitt, A. 2012, Astrophysical Journal, 750, 115

\bibitem[{Kozai(1962)}]{Kozai1962}
Kozai, Y. 1962, The Astronomical Journal, 67, 579

\bibitem[{Li \& Christou(2016)}]{Li2016}
Li, D. \& Christou, A.~A. 2016, Celestial Mechanics and Dynamical Astronomy,
  125, 133

\bibitem[{Li \& Christou(2017)}]{Li2017a}
Li, D. \& Christou, A.~A. 2017, The Astronomical Journal, 154, 209

\bibitem[{Li \& Christou(2020)}]{Li2020}
Li, D. \& Christou, A.~A. 2020, The Astronomical Journal, 159, 184

\bibitem[{Li {et~al.}(2019)Li, Mustill, \& Davies}]{Li2019}
Li, D., Mustill, A.~J., \& Davies, M.~B. 2019, Monthly Notices of the Royal
  Astronomical Society, 488, 1366

\bibitem[{Lidov(1962)}]{Lidov1962}
Lidov, M. 1962, Planetary and Space Science, 9, 719

\bibitem[{Lissauer {et~al.}(2011)Lissauer, Ragozzine, Fabrycky, Steffen, Ford,
  Jenkins, Shporer, Holman, Rowe, Quintana, Batalha, Borucki, Bryson, Caldwell,
  Carter, Ciardi, Dunham, Fortney, Gautier, Howell, Koch, Latham, Marcy,
  Morehead, \& Sasselov}]{Lissauer2011}
Lissauer, J.~J., Ragozzine, D., Fabrycky, D.~C., {et~al.} 2011, Astrophysical
  Journal, Supplement Series, 197, 8

\bibitem[{Malmberg {et~al.}(2011)Malmberg, Davies, \& Heggie}]{Malmberg2011}
Malmberg, D., Davies, M.~B., \& Heggie, D.~C. 2011, Monthly Notices of the
  Royal Astronomical Society, 411, 859

\bibitem[{Marzari \& Weidenschilling(2002)}]{Marzari2002}
Marzari, F. \& Weidenschilling, S.~J. 2002, Icarus, 156, 570

\bibitem[{McKinnon(1984)}]{McKinnon1984}
McKinnon, W.~B. 1984, Nature, 311, 355

\bibitem[{McKinnon \& Leith(1995)}]{McKinnon1995}
McKinnon, W.~B. \& Leith, A.~C. 1995, Icarus, 118, 392

\bibitem[{Morbidelli {et~al.}(2009)Morbidelli, Brasser, Tsiganis, Gomes, \&
  Levison}]{Morbidelli2009}
Morbidelli, A., Brasser, R., Tsiganis, K., Gomes, R., \& Levison, H.~F. 2009,
  Astronomy and Astrophysics, 507, 1041

\bibitem[{Morbidelli {et~al.}(2012)Morbidelli, Tsiganis, Batygin, Crida, \&
  Gomes}]{Morbidelli2012b}
Morbidelli, A., Tsiganis, K., Batygin, K., Crida, A., \& Gomes, R. 2012,
  Icarus, 219, 737

\bibitem[{Mosqueira \& Estrada(2003)}]{Mosqueira2003a}
Mosqueira, I. \& Estrada, P.~R. 2003, Icarus, 163, 232

\bibitem[{Murray \& Dermott(1999)}]{Murray1999}
Murray, C.~D. \& Dermott, S.~F. 1999, {Solar System Dynamics} (Cambridge
  University Press), 592

\bibitem[{Namouni(2010)}]{Namouni2010}
Namouni, F. 2010, The Astrophysical Journal Letters, 719, 4

\bibitem[{Nesvorn{\'{y}}(2011)}]{Nesvorny2011}
Nesvorn{\'{y}}, D. 2011, The Astrophysical Journal, 742, L22

\bibitem[{Nesvorn{\'{y}}(2015)}]{Nesvorny2015}
Nesvorn{\'{y}}, D. 2015, The Astronomical Journal, 150, 68

\bibitem[{Nesvorn{\'{y}}(2018)}]{Nesvorny2018}
Nesvorn{\'{y}}, D. 2018, Annual Review of Astronomy and Astrophysics, 56, 137

\bibitem[{Nesvorn{\'{y}} {et~al.}(2003)Nesvorn{\'{y}}, Alvarellos, Dones, \&
  Levison}]{Nesvorny2003}
Nesvorn{\'{y}}, D., Alvarellos, J. L.~A., Dones, L., \& Levison, H.~F. 2003,
  The Astronomical Journal, 126, 398

\bibitem[{Nesvorn{\'{y}} \& Morbidelli(2012)}]{Nesvorny2012}
Nesvorn{\'{y}}, D. \& Morbidelli, A. 2012, The Astronomical Journal, 144, 117

\bibitem[{Nesvorn{\'{y}} \& Vokrouhlick{\'{y}}(2016)}]{Nesvorny2016}
Nesvorn{\'{y}}, D. \& Vokrouhlick{\'{y}}, D. 2016, The Astrophysical Journal,
  825, 94

\bibitem[{Nesvorn{\'{y}} {et~al.}(2014{\natexlab{a}})Nesvorn{\'{y}},
  Vokrouhlick{\'{y}}, \& Deienno}]{Nesvorny2014}
Nesvorn{\'{y}}, D., Vokrouhlick{\'{y}}, D., \& Deienno, R. 2014{\natexlab{a}},
  The Astrophysical Journal, 784, 22

\bibitem[{Nesvorn{\'{y}} {et~al.}(2014{\natexlab{b}})Nesvorn{\'{y}},
  Vokrouhlick{\'{y}}, Deienno, \& Walsh}]{Nesvorny2014a}
Nesvorn{\'{y}}, D., Vokrouhlick{\'{y}}, D., Deienno, R., \& Walsh, K.~J.
  2014{\natexlab{b}}, The Astronomical Journal, 148, 52

\bibitem[{Nesvorn{\'{y}} {et~al.}(2007)Nesvorn{\'{y}}, Vokrouhlick{\'{y}}, \&
  Morbidelli}]{Nesvorny2007}
Nesvorn{\'{y}}, D., Vokrouhlick{\'{y}}, D., \& Morbidelli, A. 2007, The
  Astronomical Journal, 133, 1962

\bibitem[{Nesvorn{\'{y}} {et~al.}(2013)Nesvorn{\'{y}}, Vokrouhlick{\'{y}}, \&
  Morbidelli}]{Nesvorny2013}
Nesvorn{\'{y}}, D., Vokrouhlick{\'{y}}, D., \& Morbidelli, A. 2013, The
  Astrophysical Journal, 768, 45

\bibitem[{Nogueira {et~al.}(2011)Nogueira, Brasser, \& Gomes}]{Nogueira2011}
Nogueira, E., Brasser, R., \& Gomes, R. 2011, Icarus, 214, 113

\bibitem[{Petit {et~al.}(2011)Petit, Kavelaars, Gladman, Jones, Parker, {Van
  Laerhoven}, Nicholson, Mars, Rousselot, Mousis, Marsden, Bieryla, Taylor,
  Ashby, Benavidez, {Campo Bagatin}, \& Bernabeu}]{Petit2011}
Petit, J.-M., Kavelaars, J.~J., Gladman, B.~J., {et~al.} 2011, The Astronomical
  Journal, 142, 131

\bibitem[{Pfalzner {et~al.}(2005)Pfalzner, Vogel, Scharw{\"{a}}chter, \&
  Olczak}]{Pfalzner2005}
Pfalzner, S., Vogel, P., Scharw{\"{a}}chter, J., \& Olczak, C. 2005, Astronomy
  {\&} Astrophysics, 437, 967

\bibitem[{Philpott {et~al.}(2010)Philpott, Hamilton, \& Agnor}]{Philpott2010}
Philpott, C.~M., Hamilton, D.~P., \& Agnor, C.~B. 2010, Icarus, 208, 824

\bibitem[{Pollack {et~al.}(1979)Pollack, Tauber, \& Burns}]{Pollack1979}
Pollack, J.~B., Tauber, M.~E., \& Burns, J.~a. 1979, Icarus, 37, 587

\bibitem[{Rasio \& Ford(1996)}]{Rasio1996}
Rasio, F.~A. \& Ford, E.~B. 1996, Science, 274, 954

\bibitem[{Rufu \& Canup(2017)}]{Rufu2017}
Rufu, R. \& Canup, R.~M. 2017, The Astronomical Journal, 154, 208

\bibitem[{Szul{\'{a}}gyi {et~al.}(2018)Szul{\'{a}}gyi, Cilibrasi, \&
  Mayer}]{Szulagyi2018}
Szul{\'{a}}gyi, J., Cilibrasi, M., \& Mayer, L. 2018, The Astrophysical
  Journal, 868, L13

\bibitem[{Teachey \& Kipping(2018)}]{Teachey2018}
Teachey, A. \& Kipping, D.~M. 2018, Science Advances, 4, eaav1784

\bibitem[{Teachey {et~al.}(2018)Teachey, Kipping, \& Schmitt}]{Teachey2018a}
Teachey, A., Kipping, D.~M., \& Schmitt, A.~R. 2018, The Astronomical Journal,
  155, 36

\bibitem[{Tremaine {et~al.}(2009)Tremaine, Touma, \& Namouni}]{Tremaine2009}
Tremaine, S., Touma, J., \& Namouni, F. 2009, Astronomical Journal, 137, 3706

\bibitem[{Tsiganis {et~al.}(2005)Tsiganis, Gomes, Morbidelli, \&
  Levison}]{Tsiganis2005}
Tsiganis, K., Gomes, R., Morbidelli, a., \& Levison, H.~F. 2005, Nature, 435,
  459

\bibitem[{Udry \& Santos(2007)}]{Udry2007}
Udry, S. \& Santos, N.~C. 2007, Annual Review of Astronomy and Astrophysics,
  45, 397

\bibitem[{Vokrouhlick{\'{y}} {et~al.}(2008)Vokrouhlick{\'{y}}, Nesvorn{\'{y}},
  \& Levison}]{Vokrouhlicky2008}
Vokrouhlick{\'{y}}, D., Nesvorn{\'{y}}, D., \& Levison, H.~F. 2008, The
  Astronomical Journal, 136, 1463

\bibitem[{Weiss \& Marcy(2014)}]{Weiss2014}
Weiss, L.~M. \& Marcy, G.~W. 2014, The Astrophysical Journal, 783, L6

\end{thebibliography}
%
\end{document}